%% file: main.tex
\documentclass[a4paper,11pt]{article}
\pdfoutput=1
\usepackage{jcappub}
\usepackage{macro}
\usepackage{lmodern} 
\usepackage[T1]{fontenc}
\usepackage[utf8]{inputenc}
\usepackage{amsmath}
\usepackage{amssymb}
\usepackage{lineno}
\usepackage{color}
\usepackage[draft]{fixme}
\usepackage[backgroundcolor=yellow, linecolor=black]{todonotes}

\newcommand{\codeStyle}[1]{\texttt{#1}}

\title{\boldmath CRPropa 3.2 --- an advanced framework for high-energy particle propagation in extragalactic and galactic spaces}

\newcommand{\UAM}{a}
\newcommand{\Radboud}{b}
\newcommand{\Bochum}{c}
\newcommand{\RAPP}{d}
\newcommand{\GSSI}{e}
\newcommand{\ICPN}{f}
\newcommand{\RWTH}{g}
\newcommand{\MPIFR}{h}
\newcommand{\Oxford}{i}
\newcommand{\BUW}{j}
\newcommand{\Innsbruck}{k}
\newcommand{\Paris}{l}
\newcommand{\IKBU}{m}
\newcommand{\Moscow}{n}
\newcommand{\Hamburg}{o}
\newcommand{\KU}{p}
\newcommand{\VUB}{q}

\author[\UAM, \Radboud]{Rafael {Alves Batista},}
\author[\Bochum, \RAPP]{Julia {Becker Tjus},}
\author[\Bochum, \RAPP]{Julien {D\"orner},}
\author[\GSSI, \ICPN]{Andrej {Dundovic},}
\author[\Bochum, \RAPP]{Bj\"orn {Eichmann},}
\author[\Bochum, \RAPP]{Antonius {Frie},}
\author[\RWTH, \MPIFR]{Christopher {Heiter},}
\author[\Bochum, \Oxford, \RAPP]{{Mario R.} Hoerbe,}
\author[\BUW, \RAPP]{Karl-Heinz {Kampert},}
\author[\Innsbruck, \Bochum, \RAPP]{Lukas {Merten},}
\author[\RWTH]{Gero {M\"{u}ller,}}
\author[\Bochum, \RAPP, \Paris]{{Patrick} Reichherzer,}
\author[\IKBU, \Moscow]{Andrey {Saveliev},}
\author[\Bochum, \RAPP]{Leander {Schlegel},}
\author[\Hamburg]{G\"unter {Sigl},}
\author[\KU]{Arjen {van Vliet},}
\author[\VUB, \MPIFR]{Tobias Winchen}

\affiliation[\UAM]{Instituto de Física Teórica UAM-CSIC, C/ Nicolás Cabrera 13-15, 28049 Madrid, Spain}
\affiliation[\Radboud]{Radboud University Nijmegen, Department of Astrophysics/IMAPP, 6500 GL Nijmegen, The Netherlands}
\affiliation[\Bochum]{Ruhr-Universität Bochum, Universitätsstraße 150, 44801 Bochum, Germany}
\affiliation[\RAPP]{Ruhr Astroparticle and Plasma Physics Center (RAPP Center), Germany}
\affiliation[\GSSI]{Gran  Sasso  Science  Institute  (GSSI),  Viale  F.  Crispi  7,  67100  L’Aquila,  Italy}
\affiliation[\ICPN]{Institute for Cosmology and Philosophy of Nature (ICPN), Trg sv. Florijana 16, 48260 Kri\v{z}evci, Croatia}
\affiliation[\RWTH]{RWTH Aachen University, III. Physikalisches Institut A, Otto-Blumenthal-Str., 52056 Aachen, Germany}
\affiliation[\MPIFR]{Max Planck Institute for Radio Astronomy, Auf dem Huegel 69, 53121 Bonn, Germany}
\affiliation[\Oxford]{University of Oxford, Oxford Astrophysics, Denys Wilkinson Building, Keble Road, Oxford, OX1 3RH, United Kingdom}
\affiliation[\BUW]{Bergische Universität Wuppertal, Department of Physics, Gaußstrasse 20, 42119 Wuppertal, Germany}
\affiliation[\Innsbruck]{Institute for Astro- and Particle Physics, University of Innsbruck, Technikerstraße 25, 6020 Innsbruck, Austria}
\affiliation[\Paris]{IRFU, CEA, Université Paris-Saclay, F-91191 Gif-sur-Yvette, France}
\affiliation[\IKBU]{Immanuel Kant Baltic Federal University, Institute of Physics, Mathematics and Information Technology, 236016 Kaliningrad, Russia}
\affiliation[\Moscow]{Lomonosov Moscow State University,
Faculty of Computational Mathematics and Cybernetics,
119991 Moscow, Russia}
\affiliation[\Hamburg]{II. Institute for Theoretical Physics, Universit\"at Hamburg, Luruper Chaussee 149, 22761 Hamburg, Germany}
\affiliation[\KU]{Department of Physics, Khalifa University, P.O.~Box 127788, Abu Dhabi, United Arab Emirates}
\affiliation[\VUB]{Vrije Universiteit Brussel, Astrophysical Institute, Pleinlaan 2, 1050 Brussels, Belgium}

\emailAdd{crpropa@desy.de}

\abstract{The landscape of high- and ultra-high-energy astrophysics has changed in the last decade, largely due to the inflow of data collected by large-scale cosmic-ray, gamma-ray, and neutrino observatories. At the dawn of the multimessenger era, the interpretation of these observations within a consistent framework is important to elucidate the open questions in this field. CRPropa~3.2 is a Monte Carlo code for simulating the propagation of high-energy particles in the Universe.
This version represents a major leap forward,  significantly expanding the simulation framework and opening up the possibility for many more astrophysical applications. This includes, among others: efficient simulation of high-energy particles in diffusion-dominated domains, self-consistent and fast modelling of electromagnetic cascades with an extended set of channels for photon production, and studies of cosmic-ray diffusion tensors based on updated coherent and turbulent magnetic-field models. Furthermore, several technical updates and improvements are introduced with the new version, such as: enhanced interpolation, targeted emission of sources, and a new propagation algorithm (Boris push). The detailed description of all novel features is accompanied by a discussion and a selected number of example applications. 
}


\begin{document}
\maketitle
\flushbottom

 \input{01_Intro}

 \input{03_NewFeat}

 \input{04_Examples}

 \input{05_Summary}

\acknowledgments
\noindent We thank Thomas Fitoussi for implementing the TF17Field magnetic-field model, Timo Schorlepp for implementing the JF12FieldSolenoidal model, Aritra Ghosh for the first work on first-order Fermi acceleration, Matthew Weiss for first work on particle splitting,  Jens Jasche and J\"org Rachen for work on the targeting method, Alexander Korochnikin for his contribution to improve the secondary yields of electromagnetic interactions, and Simone Rossoni for testing various parts of the software.
\\

\noindent RAB is funded by the ``la Caixa'' Foundation (ID 100010434) and the European Union's Horizon~2020 research and innovation program under the Marie Skłodowska-Curie grant agreement No~847648, fellowship code LCF/BQ/PI21/11830030. RAB also acknowledges the support from the Radboud Excellence Initiative in the early stages of this work. 
LM acknowledges financial support from the Austrian Science Fund (FWF) under grant agreement number I~4144-N27.
JD, BE, KHK, LM, PR, LS, and JT acknowledge support from the Deutsche Forschungsgemeinschaft (DFG): this work was performed in the context of the DFG-funded Collaborative Research Center SFB1491 "Cosmic Interacting Matters - From Source to Signal" (JD, BE, KHK, LM, PR, LS, JT). It was further supported by the project \textit{Multi-messenger probe of Cosmic Ray Origins (MICRO)}, grant numbers TJ\,62/8-1 (PR, LS, JT) and KA\,710/5-1 (KHK). The work of AS is supported by the Russian Science Foundation under grant no.~22-11-00063. KHK and GS acknowledge support the BMBF Verbundforschung under grants 05A20PX1 and 05A20GU2. GS acknowledges support by the Deutsche Forschungsgemeinschaft (DFG, German Research Foundation) under Germany’s Excellence Strategy – EXC 2121 Quantum Universe – 390833306.
AvV acknowledges support from the European Research Council (ERC) under the European Union’s Horizon 2020 research and innovation program (Grant No.~646623). PR acknowledges support by the German Academic Exchange Service and by the RUB Research School via the \textit{Project International} funding.

\bibliography{references}
\bibliographystyle{JHEP}

\end{document}

%% file: 01_Intro.tex
\section{Introduction}

In the last decades, the high-energy Universe has been explored with increasing precision. Various instruments have been measuring the cosmic-ray composition, energy spectra, and their anisotropies from GeV to ZeV energies (see e.g.\ \cite{BeckerTjus:2020xzg,AlvesBatista:2019tlv,kampert2014} for reviews). The Fermi Large Area Telescope (Fermi-LAT) has provided us with a systematic view of the GeV gamma-ray sky~\cite{fermi_4fgl2020}; the imaging air-Cherenkov telescopes H.E.S.S., MAGIC, and VERITAS are exploring the TeV gamma-ray Universe (see \cite{hofmann2018} for a review), and IceCube detected the first astrophysical high-energy neutrinos \citep{icecube2013,icecube2014}. In order to interpret these different observations of multiple messengers within a single framework, the comprehensive theoretical modelling of the transport and interactions of cosmic rays with ambient matter and radiation is necessary. In the regime of diffusive propagation, it is a typical approach to numerically solve the cosmic-ray transport equation (see e.g.\ \citep{strong_new_1998,strong_propagation_1998,dragon,dragon2013,Kissmann2014,Kissmann2015,propPy2022}) to model propagation on galactic scales. Such diffusive approaches are also commonly used to describe cosmic-ray transport in local source environments, such as active galactic nuclei (AGN)~\cite{eichmann2012,winter2013,murase2013,reimer2019,Xavier2020,eichmann2022} or in gamma-ray bursts (GRBs)~\cite{murase2006,winter2014}. The propagation from these extragalactic sources to Earth, on the other hand, is modelled via ballistic transport, since typical extragalactic magnetic fields (EGMFs), source distances and cosmic-ray energies lead to at most a moderate number of gyrations of the charged primary particles and the diffusive limit is not reached.

The numerical cosmic-ray transport code CRPropa, developed in the early 2000s, was one of the first propagation codes to perform the transport of ultra-high-energy cosmic rays (UHECRs) across intergalactic space, thereby improving the understanding of the energy spectrum and their arrival directions, and predicting the fluxes of secondary gamma rays and neutrinos produced in photohadronic interactions ~\cite{Sigl2007PhRvD, yueksel_kistler2007PhRvD, alvesbatista2019a}. From the beginning, it enabled the simulation of one-dimensional propagation along straight lines and of three-dimensional trajectories including deflections by magnetic fields.
 
CRPropa has developed significantly since the first version, CRPropa~1.0, published in 2007~\cite{Armengaud:2006fx}. It was extended to include the treatment of photonuclear interactions that lead to the disintegration of cosmic-ray nuclei (CRPropa~2.0, \cite{Kampert:2012fi}). This way, the flux of secondary neutrinos and gamma rays from UHECR interactions could be calculated by taking into account nuclei of all masses, ranging from protons up to iron, with immediate implications for the fluxes of cosmogenic neutrinos and photons. During this time, the numerical propagation tool SimProp was released~\citep{simprop2012, aloisio2017a} and extensive  comparisons between the two codes were done for the case of one-dimensional propagation~\cite{Batista:2015mea,AlvesBatista:2019rhs}.
 
dotivated by the increasing interest of the community, the next version, CRPropa~3.0, was released, featuring a modern, modular structure~\cite{Batista:2016yrx}, which makes it easier to extend the code to other propagation environments, or to use custom input quantities. With the launch of this new version, new features were introduced, such as an interface for Galactic propagation through the so-called \emph{lensing}, necessary for the comparison of cosmic-ray propagation simulations with cosmic-ray observables including anisotropy. In addition, time-dependence was introduced in order to be able to account for cosmological evolution effects and variable sources. This way, investigations of the contribution of different extragalactic source classes to the composition and to the all-particle spectrum at ultra-high energies could be made (see e.g.\ \cite{AlvesBatista:2017shr, Eichmann2018JCAP, Eichmann2019JCAP, Rodrigues:2020pli}), and quantitative investigations of the expected anisotropy could be performed~\cite{Hackstein:2017pex, Eichmann2020JCAP}. CRPropa~3.0 represented a state-of-the-art code for extragalactic cosmic-ray propagation. The next version, CRPropa~3.1, then went a step further by adding a diffusion module that enables the propagation of particles not by simulating trajectories directly, but by solving diffusion-convection equations. The original method for simulating deterministic trajectories is limited by the resolution of turbulent grids and by the availability of computational time. The method becomes inefficient and inaccurate for propagation within the Galaxy at energies $E \lesssim 10^{17}$~eV. This new module of diffusive propagation made it possible to perform simulations of Galactic cosmic-ray sources and other diffusive environments like the Galactic termination shock \citep{Merten:2018}, external galaxies like M51 \citep{doerner2022}, and galaxy clusters \citep{AlvesBatista2021, Hussain:2022tls}. With this addition, CRPropa is the first public tool that handles both ballistic and diffusive propagation within a single framework. This is only possible because, for the diffusive propagation, the transport equation is transformed into stochastic differential equations (SDEs), which is numerically similar to the equations of motion of real particles. In this version, stochastic terms describing the diffusive process were added, with the possibility to realise diffusion tensor configurations in arbitrary background fields. This method has the advantage that particle densities are represented by pseudo-particles, whose trajectories are simulated. This makes the transport equation approach fit into the framework of ballistic particle propagation, meaning that all the already existing modules of CRPropa can be used with the diffusion module.

Here we launch the new version of CRPropa, 3.2, which significantly extends the physics scenarios that can be modelled with the code to address the theoretical challenges in astroparticle physics~\cite{AlvesBatista:2021gzc}.
We make several method improvements (section~\ref{ssec:methods}) by introducing the Boris push as an alternative algorithm for the numerical solution of the equation of motion, through the introduction of the targeting method for more efficient simulations of isotropic emission scenarios, and by improving the interpolation methods. We further add new magnetic-field models including turbulent and helical fields, as well as different parametrisations of the global magnetic field of the Milky Way, and a component for the central molecular zone. In section \ref{ssec:galactic}, the method for ensemble-averaged propagation is presented, as it was extended significantly since its first introduction in CRPropa~3.1. Changes and additions in the way radiation and particle interactions are handled are summarised in section \ref{ssec:interactions}. Here, the description of photopion production via the event generator SOPHIA~\cite{muecke2000a} was significantly rearranged by adding a user interface and by rewriting the code such that propagation in arbitrary background field configurations is now possible in a simple way, thanks to the modular structure of CRPropa. Furthermore, some photon production channels have been updated, including photodisintegration, nuclear decay, and elastic scattering, and a native implementation of electromagnetic cascades was introduced. Finally, a module for cosmic-ray acceleration via  second- or first-order Fermi mechanism was added. In section~\ref{sec:examples} we show a few examples that illustrate possible applications of the new features. The paper concludes with a short summary of the future of high- and ultra-high-energy particle propagation modelling.

%% file: 03_NewFeat.tex
\section{New Features in CRPropa 3.2}
\label{sec:newFeatures}

In this section the new features of CRPropa~3.2 are described. Methodological improvements are shown in section~\ref{ssec:methods}. New magnetic-field models, including turbulent and a regular Galactic field, are listed and explained in section~\ref{ssec:bFields}. Section~\ref{ssec:galactic} summarises the ensemble-averaged transport model and section~\ref{ssec:interactions} explains the improved interactions. The last part of this section,~\ref{ssec:acceleration}, deals with the newly-implemented acceleration modules.

\input{03a_methods}
\input{03b_bFields}

\input{03c_galacticProp}

\input{03d_interactions}

\input{03e_acceleration}

%% file: 03a_methods.tex
\subsection{Method improvement}
\label{ssec:methods}

\subsubsection{Boris Push}
\label{sssec:BorisPush}

Numerical simulations of charged particle propagation through magnetic fields with an analytically solvable equation of motion can be performed using many different algorithms. The Boris push \cite{boris1972proceedings} and the Cash-Karp method \cite{CK1990} are two common ones which generally work best for charged particle propagation in a magnetic field~\cite{Quin2013, Winkel2015}. Because they are both well-tested and perform well, we have implemented the \codeStyle{PropagationBP} module, based on the Boris push algorithm, for ballistic particle propagation, in addition to the \codeStyle{PropagationCK} module, which is based on the Cash-Karp algorithm. Both modules support adaptive step sizes and can be easily interchanged. 

Fundamentally, both algorithms differ in that Cash-Karp is not energy-conserving, whereas the Boris push conserves energy in the absence of electric fields. However, when simulating charged particles with \codeStyle{PropagationCK} without considering interactions or other energy losses, CRPropa enforces conservation of energy. Enforcing a constant particle energy leads to an adjustment of the momentum components when using the intrinsically non-energy-conserving Cash-Karp algorithm. The Boris push, on the other hand, keeps the momentum components unchanged due to its intrinsic ability to conserve energy. 

\begin{figure}[htbp]
    \centering
    \includegraphics[width=\textwidth]{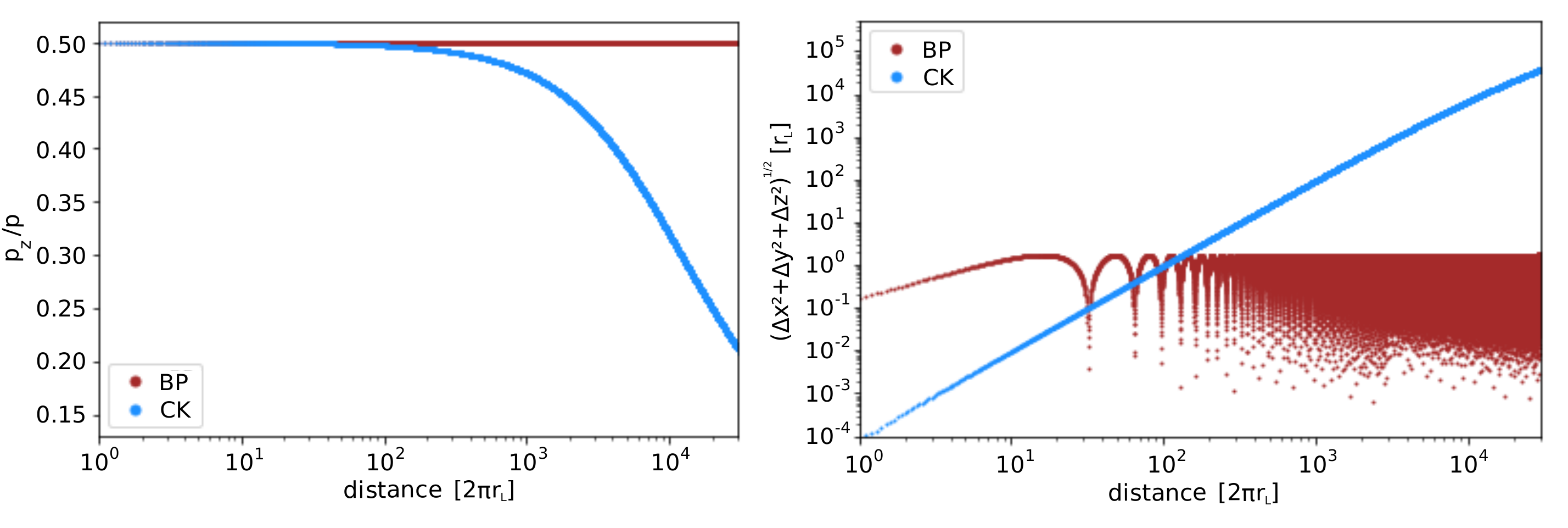}
    \caption{\emph{Left:} The momentum component along the $z$-axis as a function of time for a charged particle starting with $p_z/p = 0.5$. The simulation setup consists only of a background magnetic field in $z$-direction. Particles simulated with the \codeStyle{PropagationBP} have constant momentum along the background field, as expected. The \codeStyle{PropagationCK} module reduces the momentum along the $z$-axis for the chosen initial condition. \emph{Right:} Deviation of analytical particle trajectory in units of the gyroradius $r_\mathrm{L}$ of a particle in a magnetic field $\vec{B}=B\, \vec{e}_z$ from simulations with \codeStyle{PropagationBP} and \codeStyle{PropagationCK}. The Boris push limits the error to $2r_\mathrm{L}$, while the error increases for \codeStyle{PropagationCK} over time. It is important to note that the error depends on the step size and can be minimised with smaller steps.
    \label{fig:bp_momentum} 
    }
\end{figure}

Figure~\ref{fig:bp_momentum} shows the time evolution of the momentum components of the particles and demonstrates the conservation of these components for the Boris push, for a magnetic field oriented along the $z$-direction. In the \codeStyle{PropagationCK} module, the decrease in the parallel component of the momentum is accompanied by an increase in the perpendicular component to accommodate the error in the energy of the Cash-Karp algorithm and keep the energy constant. This behaviour has direct consequences for the particle trajectory, as can be seen in the right panel, where the deviation of the simulated trajectories with respect to the analytical case is presented. The errors in the momentum components, which accumulate along the trajectory, lead to an increasing error of the exact particle trajectory compared with the analytical solution in the \codeStyle{PropagationCK} module. The deviation arising from the Boris push method, despite the correct preservation of the momentum components, is caused by the fact that the phases are not preserved. This changes the gyro-frequency and results in the absolute distance of the particle, oscillating between 0 and $2r_\mathrm{L}$ in the simulations. However, this error is globally constrained and leads to simulated trajectories that are closer to the analytical ones, even for large times. This behaviour, together with the better temporal performance of the Boris push at comparable step sizes, favours the use of the \codeStyle{PropagationBP} module for most astrophysical applications. \codeStyle{PropagationCK}, in combination with very small step sizes and thus long simulation times, offers an advantage only if exact trajectories of single particles are required. Note that for UHECR transport through intergalactic magnetic fields (IGMFs), differences between both modules are expected to be negligible if the step sizes for both methods are small enough to sufficiently resolve the scales of the gyration motion and the turbulent fluctuations. We refer interested readers to the numerical setup described in Ref.~\cite{Reichherzer2021BP}.

Note that both single particle propagation methods (\codeStyle{PropagationCK} and \codeStyle{PropagationBP}) solve the equation of motion in comoving coordinates. Any relevant redshift ($z$) evolution of the magnetic fields needs to be explicitly included in a simulation, e.g., by using the \codeStyle{MagneticFieldEvolution} class. Using a magnetic field model constant in redshift $B(z) = B_0$, e.g., the deflection angles of the propagated cosmic rays will be proportional to the travelled comoving distance. Applying a redshift scaling to the magnetic field $B_0 \rightarrow B_0  (1+z)^{-1}$ will lead to deflection angles that are proportional to the light-travel distance.

\subsubsection{Targeting}
\label{sssec:targeting}

To obtain realistic predictions for the arrival directions of UHECRs, CRPropa can track the propagation of UHECRs from their sources to Earth through extensive structured IGMF models. In such 3D or 4D (including cosmological evolution or timing effects) simulations UHECRs are emitted isotropically from specific source positions, their trajectories are tracked through the IGMFs, and they are registered as observed if at some point they reach a relatively small target, the observer sphere. If the trajectory of the UHECR never actually hits the observer sphere the simulation is disregarded. Therefore, it can be very computationally intensive to obtain enough statistics for reliable predictions of arrival-direction distributions in such scenarios. To remedy this problem, we have developed a `targeting' method~\citep{Jasche:2019sog}. In this targeting mechanism the UHECRs are not emitted isotropically from their sources, but an optimal directional emission distribution is found to maximise the number of hits. In this process the particles are each assigned a certain weighting factor which ensures that the resulting arrival-direction distribution remains unbiased. The same arrival-direction distributions are, therefore, obtained as if the UHECRs were emitted isotropically from their sources. This method can lead to speedups (see below) of several orders of magnitude in the number of particles hitting the observer sphere, compared with the case of isotropic emission from the sources, depending on the specific setup of the simulation.

In this targeting method the UHECRs are emitted from their sources following a probability distribution for random emission directions around a preferred direction. This probability distribution is given by the von~Mises-Fischer (vMF) distribution on a sphere $\Pi(\vec{x})$ for $\vec{x}$, a unit vector of a random direction on the sphere,
with mean direction $\vec{\mu}$, the Cartesian unit vector corresponding to the preferred direction of emission, and concentration parameter $\kappa$, which controls the width of the distribution. The values for $\vec{\mu}$ and $\kappa$ can be optimised by learning from previous simulations, whether or not they hit the observer sphere.

The initial optimal emission direction $\vec{\mu}_0$ is simply chosen to point straight from the source to the observer.
The initial value for $\kappa$ is chosen by requiring, in the case of straight-line propagation, that a fraction $p$ of emitted particles hits the observer sphere. Therefore, $\kappa$ is initially set to:
\begin{equation}
\label{eq:estimate_kappa}
\kappa_0 = \frac{\mathrm{ln}(1-p)}{a_0-1}\, ,
\end{equation}
with $a_0$ given by the apparent detector size; $a_0 = \cos (\mathrm{arctan}(s/D))$, with $s$ denoting the radius of the observer sphere and $D$ the distance between the source and the detector. The choice of $p$ determines how many particles initially will be emitted in the direction of the observer sphere. A larger value of $p$ is likely to increase the yield of detected particles but limits the exploration of other possible directions that might lead to hits of the observer sphere as well. Generally speaking, for lower rigidities and stronger magnetic fields with larger coherence lengths, a smaller $p$ should be chosen. Typical values may range from $p=0.1$ when large deflections are expected, to $p=0.9$ when only small deflections are expected. No matter the choice of $p$, in the large-sample limit the algorithm will converge to the correct result, and will just do it faster in case of a better choice of $p$. 

At the start of the propagation each particle is assigned a weight $\omega_n$ to account for the emission according to the vMF distribution while the resulting arrival-direction distribution should correspond to an isotropic emission from the sources. This weight depends on the emission direction of the particle and is given by:
\begin{equation}
\omega_n \;=\; \frac{1}{4\pi} \, \frac{1}{\Pi(\vec{x})} \, .
\end{equation}
The weights $\omega_n$ guarantee that the correct arrival-direction distribution is obtained in the large-sample limit.

The optimisation of $\vec{\mu}$ and $\kappa$ can be obtained by running the simulation in several batches of $N_{\mathrm{batch}}$ particles. In the first batch the emission directions will be drawn from a vMF distribution with $\vec{\mu}_0$ and $\kappa_0$. In the following batches the values of $\vec{\mu}$ and $\kappa$ will be adjusted according to which particles of the previous batch hit the observer sphere. The updated preferred emission direction is estimated as:
\begin{equation}
\vec{\mu} = \dfrac{\sum\limits_{n=0}^{N_{\mathrm{hit}}} \vec{x}_n \, \omega_n  }{ \left|\sum\limits_{n=0}^{N_{\mathrm{hit}}} \vec{x}_n \, \omega_n\right|} \, ,
\end{equation}
with $\vec{x}_n$ the successful emission directions, $\omega_n$ the corresponding importance weights and $N_{\mathrm{hit}}$ the number of particles that hit the observer sphere. The parameter $\kappa$ is updated using Eqn.~\ref{eq:estimate_kappa} with $a_0$ updated to a new value using:
\begin{equation}
a = \dfrac{\sum\limits_{n=0}^{N_{\mathrm{hit}}} \left(\vec{\mu}^T \vec{x}_n \right)^2 \, \omega_n  }{\sum\limits_{n=0}^{N_{\mathrm{hit}}}  \omega_n}\, .
\end{equation}
For the mathematical derivation of the validity of this approach see~\citep{Jasche:2019sog}.

\begin{figure}[htbp]
    \centering
    \includegraphics[width=0.7\linewidth]{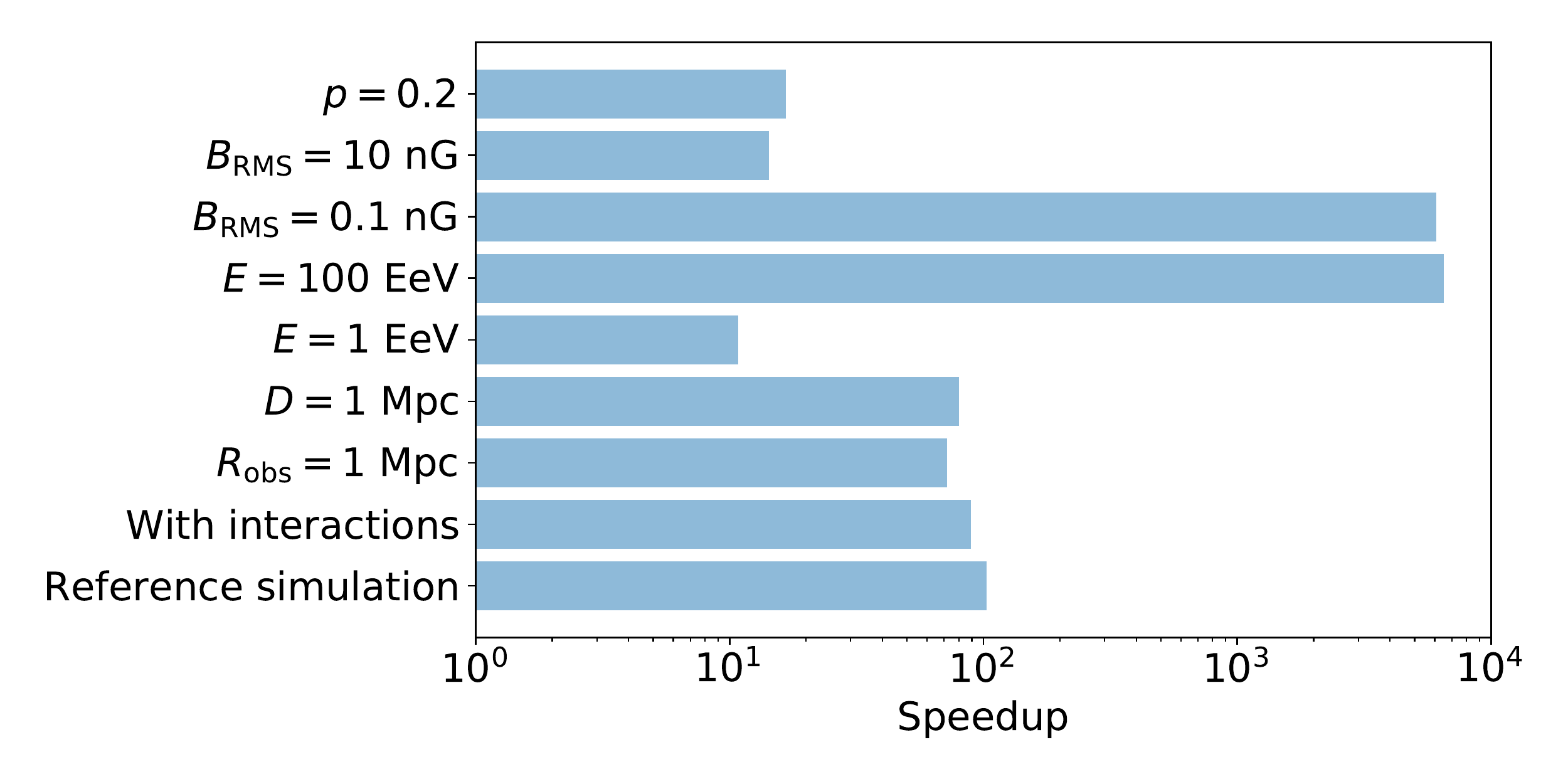}
    \caption{Speedup of CRPropa by using targeted emission \citep{Jasche:2019sog}. The mean direction and concentration parameter of the vMF distribution are given by $\vec{\mu}_0$ and $\kappa_0$ without any optimisation. The reference simulation has the following setup: protons with an energy of $E = 10$~EeV are emitted from  source at a distance of $D = 10$~Mpc from a spherical observer with a radius of $R_{\mathrm{obs}} = 0.1$~Mpc. The magnetic field is a Kolmogorov-type turbulent magnetic field with $B_{\mathrm{rms}} = 1$~nG, the hit probability is set to $p = 0.1$ and no interactions are included. For the other scenarios, one parameter of the reference simulation is changed (as indicated in the figure) while all other parameters are the same as in the reference simulation.
    \label{fig:TargetingSpeedup} 
    }
\end{figure}

An estimate for the speedup that can be achieved with this method is given in Fig.~\ref{fig:TargetingSpeedup}.
The speedup in this case is defined as the ratio of simulation times per observed hit for isotropic emission over targeted emission, $\frac{t_{\mathrm{iso}}/N_{\mathrm{iso}}}{t_{\mathrm{tar}}/N_{\mathrm{tar}}}$, with $t_{\mathrm{iso}}$ ($t_{\mathrm{tar}}$) and $N_{\mathrm{iso}}$ ($N_{\mathrm{tar}}$) the time the simulation took, and the number of hits detected at the observer for isotropic (targeted) emission.
This figure shows conservative estimates of the speedup as the mean direction and concentration parameter of the vMF distribution were not optimised but were fixed to their initial values $\vec{\mu}_0$ and $\kappa_0$. The speedup depends strongly on the specifics of the simulation setup with, generally speaking, larger speedups for scenarios with weaker deflections in magnetic fields. For the scenarios tested here, speedups between one and four orders of magnitude were obtained. 

\subsubsection{Enhanced Interpolation}
\label{sssec:interpolation}

To obtain a continuous description from a given set of discrete data, the use of an interpolation algorithm becomes inevitable. Such an interpolation algorithm is, for example, used in CRPropa to obtain the magnetic-field strength at an arbitrary position from a grid of magnetic-field data to compute the particle trajectory. In principle, each interpolation algorithm provides information that is not based on the given grid of data, but the underlying assumptions on the correlation between the grid points. To avoid any assumptions on those correlations, the so-called \emph{nearest-neighbour} (NN) method can be employed, where the data from the closest grid point with respect to the current spatial position is used without taking any additional grid points into account. Supposing a linear correlation of the data of neighbouring grid points yields the \emph{trilinear} (TL) method --- which has been the only available interpolation routine in previous versions of CRPropa, and is still the default. If the information of the eight closest grid points is not sufficient, as also their neighbours affect the outcome, it can become necessary to use a \emph{tricubic} (TC) method~\cite{LekienMarsden2005}, where the information of 64 grid points are included. 

\begin{figure}[htb]
    \centering
    \includegraphics[width=\textwidth]{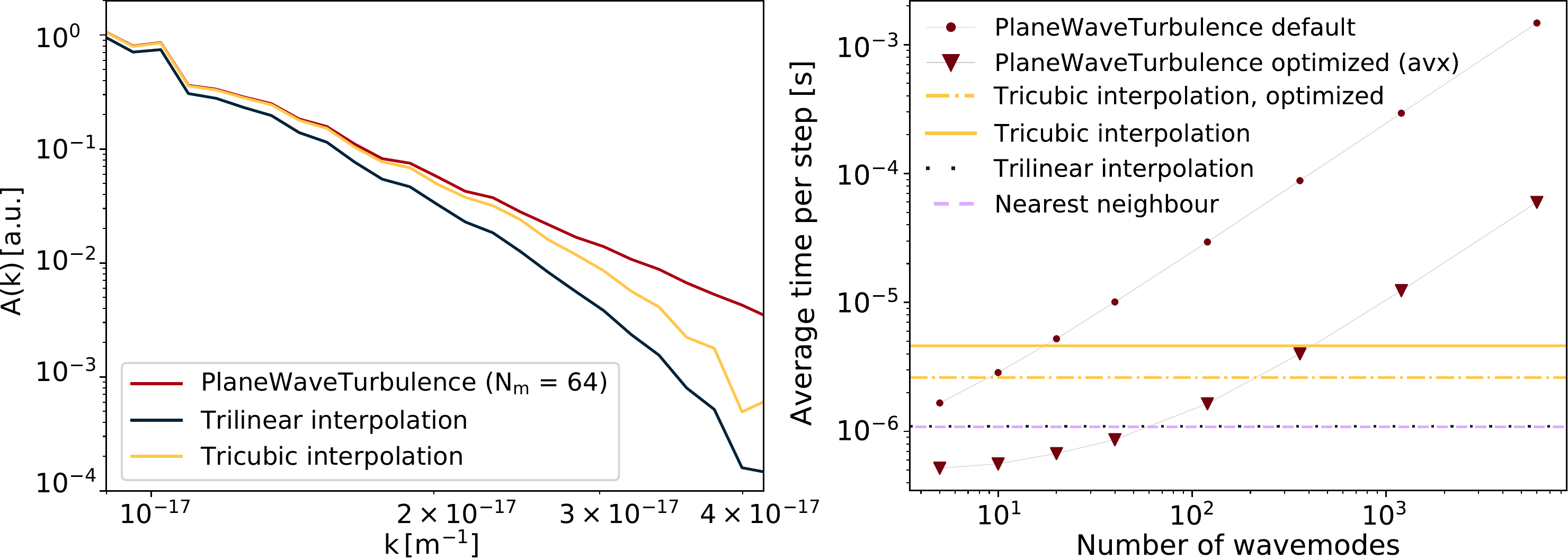}
    \caption{\emph{Left:} Kolmogorov spectrum of magnetic turbulence in the $k$-space as generated by plane-wave turbulence (see section~\ref{sssec:turbulentFields} for more details) with and without a subsequent interpolation of the data. \emph{Right:} Performance of various magnetic-field methods in test simulations. The grid-based interpolation methods do not have a parameter that affects their runtime, so they are drawn as horizontal lines. Data for plane-wave turbulence is shown for different numbers of wave-modes. 
    \label{fig:turbSpec_perf} 
    }
\end{figure}

It has been shown in Ref.~\cite{Schlegel+2020} that the average properties of a discretised, turbulent magnetic field can change significantly by the interpolation, especially when the TL algorithm is applied to a low-resolution grid. In this case, the root mean square of the field strength ($B_{\rm rms}$) becomes smaller (for TL or TC), such that the divergence of the interpolated field no longer vanishes and the turbulent spectrum lacks wave modes with large wave numbers, causing a steepening of the spectral behaviour (c.f.\ Fig.~\ref{fig:turbSpec_perf}, left). 
By construction, the NN method yields the correct $B_{\rm rms}$ value and requires smaller computation time (c.f.\ Fig.~\ref{fig:turbSpec_perf}, right), however, it also introduces discontinuities between the individual grid cells. 
TL interpolation is only slightly slower but needs a smaller grid spacing to provide a high accuracy of the magnetic-field properties. TC interpolation takes about two (four) times longer with (without) ``single instruction multiple data'' (SIMD) processor instructions, but best reproduces the turbulent magnetic-field structure and its average properties. The impact of the interpolation error on the trajectory of a particle depends strongly on its rigidity and the details of the turbulent magnetic field. The interested reader is referred to Ref.~\cite{Schlegel+2020} for more information on its impact in the resonant scattering regime of diffusive motion by a purely turbulent magnetic field. 

In general, there is no optimal interpolation method as each one has its advantages and disadvantages. The decision on which method to use to solve a particular problem is mostly driven by a compromise between performance and accuracy. Besides the TL interpolation, the NN and TC interpolation methods are now also available in CRPropa. In principle, we recommend the usage of a small grid spacing and to utilise the available storage to a priori minimise this impact. Alternatively, the continuous, turbulent magnetic field (introduced as plane-wave turbulence in section~\ref{sssec:turbulentFields}), which does not require any interpolation, can also be used. 

\subsubsection{Galactic Gas Densities}
\label{sssec:massDensities}

CRPropa now includes an interface for Galactic gas distributions. They can be used as hadronic targets for upcoming interaction modules or to model the Galactic source distribution based on the distribution of gas in the Milky Way.

The implementation is similar to that of magnetic fields, allowing the combination of two or more models into one, making the addition of user-defined models easier. CRPropa~3.2 provides several models of the distributions of different states of interstellar hydrogen: atomic (H-I), ionised (H-II), and molecular (H$_2$). All models contain three boolean flags which turn each component on and off. All modules of the \codeStyle{Density} class give the number density $n = n_{\textrm{H-I}} + n_{\textrm{H-II}} + n_{\rm{H_2}}$ (\codeStyle{getDensity}) and the nucleon number density $n_{\textrm{nucl}} = n_{\textrm{H-I}}+n_{\textrm{H-II}} + 2n_{\rm{H_2}}$ (\codeStyle{getNucleonDensity}).

Implemented are the H-II model of Cordes \textit{et al.}~\cite{Cordes1991} and the H-I and H$_2$ models of Nakanishi \textit{et al.}~\cite{Nakanishi2003, Nakanishi2006}. Additionally, a combined model for H-I, H-II, and H$_2$ by Ferrière \textit{et al.}~is provided, using the description of \cite{Ferriere2007} in the inner Galaxy ($r_\mathrm{gc} < 3 ~ \mathrm{kpc}$) and \cite{Ferriere1998} in the outer Galaxy ($r_\mathrm{gc} \geq 3~\mathrm{kpc}$), where $r_\mathrm{gc}$ is the radial distance from the Galactic centre.


%% file: 03b_bFields.tex
\subsection{Magnetic fields}
\label{ssec:bFields}

\subsubsection{Turbulent Magnetic Fields}
\label{sssec:turbulentFields}

Since its inception, CRPropa was mainly used to study high-energy cosmic rays in the intergalactic context. As of version 3.0, it also provides the technique of cosmic-ray lensing inherited from the PARSEC software\,\cite{Bretz:2013oka} to take into account deflections in the arrival directions caused by Galactic magnetic fields. However, with the emergence of version 3.1\,\cite{Merten:2017mgk}, the code has gained traction in the research of Galactic cosmic-ray diffusion\,\cite{Reichherzer:2019dmb,Dundovic:2020sim,Reichherzer:2021a}. These studies focus on cosmic rays whose gyroradii, \(r_\text{L}\), tend to be comparable or smaller than the correlation length, \(L_B\), of the turbulent magnetic field, in contrast to the regime of high-energy cosmic rays which propagate through intergalactic turbulent magnetic fields where \(r_\text{L} \gtrsim L_B\).  In the latter case, the diffusion coefficients are easier to model (e.g.\,\cite{Aloisio:2004jda}), while in the former case the resonance between the cosmic-ray gyration and turbulent wave-modes dictates the behaviour of the cosmic-ray transport so that the involved turbulent field requires a detailed description and a more sophisticated approach in its numerical implementation. 
Therefore, the requirements for propagating CRs in a turbulent magnetic field in the regime of \(r_\text{L} \lesssim L_B\) resulted in several improvements of the code.%

Firstly, the turbulent magnetic field featured in CRPropa~3.0 has been completely redesigned while still allowing the use of the previous interface (e.g., \codeStyle{initTurbulence}), albeit it is marked as obsolete. Previously, the turbulent field spectrum was modelled as a power law ranging from the largest scales of turbulence, \(L_\mathrm{max}\), to the smallest, \(L_\mathrm{min}\). This so-called \textit{cut-off} model of the spectrum has been kept, but renamed to \codeStyle{SimpleGridTurbulence}. The new implementation of turbulent fields now allows for simulations going beyond the simple \emph{cut-off} model, e.g., a \emph{bendover} scale can be defined leading to a smooth power spectrum. Furthermore, different spectral indices for the injection and cascade part of the power spectrum can be used.
Secondly, to avoid the problems associated with interpolating grid-based data (see section~\ref{sssec:interpolation}), one might consider using a continuous model for the magnetic field instead, i.e.\ by defining the magnetic field through an equation that can be evaluated at each point in space.

For a turbulent magnetic field, such an approach has been described in various works~\cite{GJ99, TD13}. The approach currently implemented in CRPropa uses an inverse discrete Fourier transform (IDFT) to compute the magnetic field as a superposition of planar waves of different amplitudes, wave numbers, and directions, and that it can do so efficiently (even for large numbers of wave modes) by evaluating the field on a whole grid of points at once. This grid is then stored and queried via an interpolation routine.
However, this process of summation over a set of wave modes can be performed manually, without the use of an IDFT, and for arbitrary points in space. Doing so also removes the requirement to have all wave modes be present on a rigid grid, and thus allows them to be distributed in a way that optimises the physical qualities of the resulting field while minimising the number of wave modes needed to represent it. This turns out to be very important, since if this computation is now to be performed at runtime, for each individual step, instead of taking up time once during setup, performance becomes a critical aspect which severely limits the number of wave modes that can realistically be used.

The \codeStyle{PlaneWaveTurbulence} magnetic-field class represents a practical implementation of this method, as described by \cite{TD13}, and it comes with an optimised evaluation routine that matched the performance of the trilinear interpolation at around 100 wave modes on our test setup (c.f.\ Fig.~\ref{fig:turbSpec_perf}).

\subsubsection{Helical Magnetic Fields}
\label{sssec:helical}

There has been a growing interest in helical magnetic fields, given their impact on the propagation of charged particles. They can lead to unique morphological signatures in gamma-ray-induced electromagnetic cascades~\cite{Long:2015bda, AlvesBatista:2016urk} as well as in the large-scale distribution of UHECRs~\cite{AlvesBatista:2018owq}. In addition, helical magnetic fields are also thought to play a role in source physics by being present in the inner jet zone of blazars (see, for example, \cite{Myserlis:2018see}).

The magnetic helicity $H_{B}$ is given by the expression:
\begin{equation}
H_{B} = \int h(\vec{r}) \, {\rm d}^{3}r = \int \vec{A}(\vec{r}) \cdot \vec{B}(\vec{r}) \, {\rm d}^{3}r \,,
\end{equation}
where $\vec{A}$ is the vector potential, $\vec{B}$ is the magnetic-field strength and $h = \vec{A} \cdot \vec{B}$ denotes the local magnetic helicity. $H_{B}$ is an important quantity in electrodynamics and magnetohydrodynamics (MHD) as it is gauge-invariant, conserved for ideal MHD, and plays a significant role in plasma relaxation~\cite{PhysRevLett.33.1139}.

In addition, there is an interesting relation between the energy $E_{B}$ and the helicity of a magnetic field, the former being given by:
\begin{equation}
E_{B} = \frac{1}{2\mu_{0}}\int \vec{B}^{2}(\vec{r}) {\rm d}^{3}r\,,
\end{equation}
where $\mu_{0}$ is the vacuum permeability. Rewriting both $H_{B}$ and $E_{B}$ in terms of their spectra $\tilde{H}_{B}(k)$ and $\tilde{E}_{B}(k)$, respectively, i.e.\ $E_{B} = \int \tilde{E}_{B}(k) \, {\rm d}\ln k \,,\, H_{B} = \int \tilde{H}_{B}(k) \, {\rm d}\ln k$, one can show \cite{Brandenburg20051} that the relation $\frac{k}{2 \mu_{0}} \left| \tilde{H}_{B}(k) \right| \leq \tilde{E}_{B}(k)$ holds, i.e.~that the maximum magnitude of the magnetic helicity spectrum for a given wave number $k$ is limited by the corresponding energy spectrum value. This may be reformulated as:
\begin{equation}
\tilde{H}_{B}(k) = f_{\rm H} \frac{2 \mu_{0}}{k} \tilde{E}_{B}(k)\,.
\end{equation}
In general it is $-1 \leq f_{\rm H} \leq 1$, however in CRPropa we limit the allowed values to $f_{\rm H} = - 1$ and $f_{\rm H} = + 1$, i.e.~to the negative and positive maximum magnetic helicity, respectively. This fraction $f_{\rm H}$ of the maximum helicity value (which in general may depend on $k$) serves as the central input value for the specification of the magnetic helicity in CRPropa.

We consider two kinds of magnetic helicity setups: a single-mode configuration~\cite{Long:2015bda,AlvesBatista:2018owq} and a stochastic magnetic field with a given power-law energy spectrum~\cite{AlvesBatista:2016urk}.

For the former, the magnetic field is given by:
\begin{equation} \label{BdefAmplitude1}
\vec{B}(\vec{r}) = B_{0} \left\{\cos\left[ \vec{k} \cdot \left( \vec{r} - \vec{r}_{0} \right) \right] \vec{e}_{1} + \sigma \sin\left[ \vec{k} \cdot \left( \vec{r} - \vec{r}_{0} \right) \right] \vec{e}_{2}\right\}\,,
\end{equation}
where $B_{0}$ is the maximum amplitude of the magnetic field, $\vec{e}_{1}$ and $\vec{e}_{2}$ are two non-parallel vectors of norm one and $\vec{k}$ is a wave vector with $\vec{k} \parallel \vec{e}_{2} \times \vec{e}_{1}$. Equation~\ref{BdefAmplitude1} may also be expressed in terms of the mean magnetic-field amplitude $B_{\rm rms}$ as: 
\begin{equation} \label{BdefAmplitude2}
\vec{B}(\vec{r}) = \sqrt{\frac{2}{1 + \sigma^{2}}} B_{\rm rms} \left\{\cos\left[ \vec{k} \cdot \left( \vec{r} - \vec{r}_{0} \right) \right] \vec{e}_{1} + \sigma \sin\left[ \vec{k} \cdot \left( \vec{r} - \vec{r}_{0} \right) \right] \vec{e}_{2}\right\}\,.
\end{equation}
For both representations of the single-mode case $\sigma$ describes the polarisation (linear, spherical, elliptic) of the magnetic field. In order to have a simple relation for that, the pair of vectors $(\vec{e}_{1},\vec{e}_{2})$ is restricted to the case of being orthogonal, $\vec{e}_{1} \perp \vec{e}_{2}$, as then $\sigma=0$ and $\sigma = \pm 1$ correspond to linear and circular polarisation, respectively, while all other values of $\sigma$, which in general is restricted to $-1 \le \sigma \le 1$, give the general elliptic case. 

The second helical magnetic-field configuration considers a stochastic magnetic field with $\left< \tilde{E}_{B}(k) \right> \propto k^{\alpha}$, which is created by sampling the Fourier-transformed magnetic field $\tilde{B}(\vec{k})$ in $k$-space from a normal distribution with zero mean and a standard deviation given by $2(2\pi/k)^{3} \tilde{E}_{B}(k)$, and by subsequently applying the IDFT in order to obtain the magnetic field in real space.

\subsubsection{New Galactic Magnetic Field Models}
\label{ssec:gmf}

The Galactic magnetic field (GMF) is not only essential for ensemble-averaged Galactic cosmic-ray propagation (see section \ref{ssec:galactic}) but has a significant effect on UHECRs, too. In this section, a short summary of the newly available Galactic magnetic field models is given.

\paragraph{Central Galactic Zone.}
\label{sssec:GalacCenter}

For global GMF models like the one implemented in the Jansson \& Farrar (\codeStyle{JF12Field}) model~\cite{Jansson2012a, Jansson2012b} and also the improved \codeStyle{JF12FieldSolenoidal} (see below), a superposition is necessary with a magnetic-field model near the Galactic Centre. This combination is needed because the global models do not include in-plane components near the centre.  
For this purpose, an additional diffuse inter-cloud component was proposed by Guenduez et al.~\cite{Guenduez2020}.
The GBFD20 model provides a diffusive component which can be deployed for the superposition and several detailed structures that could be utilised for detailed simulations of processes in the Galactic Centre region. It covers in detail 13 molecular clouds, 7 non-thermal filaments, and the radio arc. It is implemented as \codeStyle{CMZField} in CRPropa.

\paragraph{Galactic halo models.}
\label{ssssec:TF17}

The structure of the Galactic halo field was examined in \cite{Terral2017} and several axi- and bisymmetric halo magnetic fields were combined with different descriptions of the disc field. Fitting the resulting synthetic Faraday depth maps to observational data based on Markov Chain Monte Carlo (MCMC) showed a significantly better fit to the data by bisymmetric models. The two best fitting halo models C0 and C1 in combination with disc models Ad1, Bd1, and Dd1 are now available as \codeStyle{TF17Field} in CRPropa and they can be freely combined. %
The model does not include an explicit description of the turbulent component of the field. The interested reader is referred to appendix~A of Ref.~\cite{Terral2017} as a starting point to include such a component in the simulations (see also section \ref{sssec:turbulentFields}).

\paragraph{Fully solenoidal update of JF12 field.}
\label{ssssec:JF12Sol}

The original model of the Galactic magnetic field of Refs.\,\cite{Jansson2012a, Jansson2012b} includes regions in which the magnetic divergence constraint is violated. This problem was addressed in \cite{Kleimann2019}, where additionally the x-shaped poloidal halo field was smoothed in the Galactic disc. The parabolic x-shape field lines that are inspired by models of Ferrière and Terral~\cite{Ferriere2014}, remove the current sheet in the Galactic disc and simultaneously improve the field line tracking, needed for the \codeStyle{DiffusionSDE} module at low Galactic heights $|z|\lesssim 300$~pc. However, these changes to the underlying model have not been refitted to the available data. Therefore, the implemented default parameters should be carefully checked by the user.

\paragraph{JF12 field update based on Planck observations.}
\label{ssssec:Planck}

The Planck collaboration updated several GMF models based on their observations~\cite{2016A&A...596A.103P}. In this work only the fitted values of the free parameters were updated while the model itself remained the same. These updated parameters for the \codeStyle{JF12Field} have been made available with the new magnetic-field class \codeStyle{PlanckJF12bField}. The main difference with respect to the original version is a reduced random component and some changes in the strength of the magnetic field of the spiral arms.

\paragraph{Archimedean spiral.}
\label{ssssec:ArchSpiral}

It is likely that the structure of the Galactic magnetic field reaches out beyond the observationally confirmed $\sim{20}$~kpc. A simple description of this halo field can be realised by an Archimedean spiral as already discussed, e.g.,\ in \cite{1987ApJ...312..170J}:
\begin{align}
\frac{\vec{B}}{B_0} = \underbrace{\mathrm{sign}{\left(\theta - \frac{\pi}{2}\right)}}_{\text{change direction at z=0}}\left(\frac{r_{\mathrm{ref}}^2}{r^2}\vec{e}_r - 
\frac{\Omega r_{\mathrm{ref}}^2 \sin(\theta)}{r v_w} \vec{e}_\phi \right) . \quad \label{eq:ArchmedeanSpiral}
\end{align}
Since observational constraints on the properties of the halo field are rare, the rotational speed $\Omega r_\mathrm{ref} \sin(\theta)$ at the reference level $r_{\mathrm{ref}}$, the wind speed $v_\mathrm{w}$, and the magnetic-field strength $B_0$ do not have any default parameters. Values for an application on cosmic rays from the Galactic termination shock can be found in \cite{Merten:2018}. Alternatively, the GMF models (see sections \ref{ssssec:TF17} and \ref{ssssec:JF12Sol}) could be continued with minimal changes to the existing modules.

%% file: 03c_galacticProp.tex
\subsection{Ensemble-Averaged Propagation}
\label{ssec:galactic}

In some situations the single-particle approach --- solving the equation of motion for each cosmic-ray individually --- is neither useful nor computationally feasible. This is usually the case when the particle transport becomes diffusive, e.g.\, in the case of Galactic cosmic rays with energies $\lesssim 3$~PeV in magnetic fields with an average strength of ${\sim}\mu\mathrm{G}$. Here, an averaged description of the complete particle ensemble (or phase-space distribution) is the method of choice. In CRPropa~3.1 this ensemble-averaged approach was introduced with the module \codeStyle{DiffusionSDE}, which has now been extended to include advective transport and corresponding adiabatic effects.

\subsubsection{Transport Equation}
\label{sssec:galactic}

In most cases, the dynamics of the cosmic-ray density, $n$, can be described by a transport equation where the information of individual trajectories is no longer necessary. In CRPropa, the so-called Parker transport equation, a Fokker-Planck partial differential equation, is used to describe the time evolution of the isotropic particle density $n$ :
\begin{align}
    \frac{\partial n}{\partial t} + \vec{u}\cdot\nabla n &= \nabla\cdot(\hat{\kappa}\nabla n) + \frac{1}{p^2}\frac{\partial}{\partial p}\left(p^2\kappa_{p}\frac{\partial n}{\partial p}\right) + \frac{p}{3}(\nabla\cdot \vec{u})\frac{\partial n}{\partial p} + S \:. \label{eq:ParkerTransport}
\end{align}
Here, advection is characterised by the wind speed $\vec{u}$ and spatial and momentum diffusion are given by $\hat{\kappa}$ and $\kappa_{p}$, respectively. 

Conventionally, Eqn.\ (\ref{eq:ParkerTransport}) is discretised in real and momentum space. For Galactic applications, it is solved by programs like GALPROP \citep{strong_propagation_1998}, DRAGON \citep{dragon, dragon2013, evoli2017a, evoli2018a}, or PICARD \citep{Kissmann2014, Kissmann2015}. Another ansatz is the transformation of Eqn.\ \ref{eq:ParkerTransport} into a set of equivalent stochastic differential equations (SDEs) (see, e.g. \citep{Gardiner2009} for the mathematical background and \cite{Kopp:2012, Muraishi2009, Miyake2015, Busching2011, Effenberger2012} for applications to CR transport). In CRPropa this feature has been introduced in version 3.1 \citep{Merten:2017mgk} and extended to selected advective fields in \citep{Merten:2018}. Neglecting momentum diffusion, which is suppressed by a factor of ${\sim}(v_\mathrm{A}/c)^2$ compared to spatial diffusion, the corresponding set of equations reads:
\begin{align}
	\mathrm{d} \vec{x} &= \vec{u}\,\mathrm{d}t + \hat{D}\,\mathrm{d}\vec{\omega} \: ,\label{eq:EMspatial}\\[8pt]
	\mathrm{d} p &= -\frac{p}{3} \nabla\cdot\vec{u} \,\mathrm{d}t \: , \label{eq:EMMomentum}
\end{align}
where $\hat{D}\hat{D}^\dagger=\hat{\kappa}+\hat{\kappa}^\mathrm{t}$ and $\mathrm{d}\vec{\omega}=\vec{\eta}\sqrt{\mathrm{d}t}$ is a Wiener-process with $\eta_i$ being normally distributed random variables. Within the local frame of the magnetic background field line $\{\vec{e}_\mathrm{t}, \vec{e}_\mathrm{n}, \vec{e}_\mathrm{b}\}$ --- where $\vec{e}_\mathrm{t}=\vec{B}_0/B_0$ is the tangential direction completing with the normal and binormal directions the local trihedron --- the diffusion tensor $\hat{\kappa}$ becomes diagonal. Whenever the curvature radius of the magnetic field lines is large enough, the two perpendicular directions ($\vec{e}_\mathrm{n}, \vec{e}_\mathrm{b}$) are degenerated which reduces the diffusion tensor to $\hat{\kappa}=\mathrm{diag}(\kappa_\parallel, \kappa_\perp, \kappa_\perp)$.
Based on these assumptions, the SDEs can be solved by the Euler-Maruyama-scheme \citep{CAM96}:
\begin{align}
\vec{x}_{n+1} - \vec{x}_n &= \left(u_x \vec{e}_x +  u_y \vec{e}_y +  u_z \vec{e}_z \right) h + \left(\sqrt{2\kappa_{\parallel}}\,\eta_\parallel\, \vec{e}_\parallel + 
\sqrt{2\kappa_{\perp}}\,\eta_{\perp, 1}\, \vec{e}_{\perp, 1} +
\sqrt{2\kappa_{\perp}}\,\eta_{\perp, 2}\, \vec{e}_{\perp, 2}\right) \sqrt{h} \: , \label{eq:EMScheme} 
\end{align}
where $h_i=t_{i+1}-t_i$ is the integration time step. The parallel ($\vec{e}_\parallel$) and perpendicular ($\vec{e}_{\perp, 1/2}$) diffusion directions form the trihedron of the local magnetic field line averaged over the propagation step (see \cite{Merten:2019phd} for details). Since the direction of the diffusion is extracted on-the-fly from the magnetic field, in principle, arbitrary diffusion models can be realised, as long as the magnetic-field model is continuously differentiable. This makes the approach much more flexible compared to other diffusive transport codes, where a change in the background magnetic field requires an often complicated re-calculation of the space-dependent diffusion tensor. As the influence of the turbulent magnetic field is encoded in the diffusion coefficient, turbulent magnetic-field components must not be included in the magnetic field used in the module \codeStyle{DiffusionSDE}. 

Equations \ref{eq:EMspatial} and \ref{eq:EMMomentum} can be seen as the equations of motion of phase-space elements (or pseudo-particles) and they are technically not very different from their real-particle counterparts. Therefore, the SDE ansatz fits perfectly into the CRPropa structure and can be used together with many of the available modules. Example applications of this technique are presented in section \ref{ssec:GalPropExample} and can be found in \cite{Merten:2017mgk, Merten:2018}.

Interpreting the outcome of a simulation based on the module \codeStyle{DiffusionSDE} needs to be done with care as trajectories of individual pseudo-particles are almost meaningless. The user has to keep in mind that only by averaging over a sufficiently large number of pseudo-particles a reasonable approximation of the transport equation will be obtained.

In summary, the ensemble-averaged description --- based on the modules \codeStyle{DiffusionSDE}, \codeStyle{AdiabaticCooling}, and the \codeStyle{AdvectionField} class --- allows for cosmic-ray simulations that are dominated by diffusive transport within the CRPropa framework.

\subsubsection{Advection Fields}
\label{sssec:advection}

The direction of the diffusion process is extracted from the magnetic-field class. Based on the same structure the \codeStyle{AdvectionField} class was implemented. To properly work with the diffusion module, this class provides the advection field ($\vec{u}$) and separately the divergence of the advection field ($\nabla\cdot\vec{u}$). Similarly to the magnetic fields, the advection fields can be freely combined within the \codeStyle{AdvectionFieldList}.

The divergence of the advection field is not needed for the \codeStyle{DiffusionSDE} itself, but only in the \codeStyle{AdiabaticCooling} (see section~\ref{sssec:adibaticcooling}) and should therefore always be implemented in any advection model.

\subsubsection{Adiabatic Energy Changes}
\label{sssec:adibaticcooling}

When the advection field has a non-negligible divergence $\nabla\cdot\vec{u}\neq 0$, the particle ensemble will be compressed or expanded leading to adiabatic heating or cooling, respectively. These adiabatic energy changes occur in several astrophysical situations, such as e.g.\ in the vicinity of a shock, where CRs can gain a significant amount of energy by adiabatic heating in addition to diffuse shock acceleration (see e.g.\ \cite{Merten:2018}). 

Similarly to other processes and interactions, adiabatic energy changes are implemented in a separate module called \codeStyle{AdiabaticCooling} which takes the advection field as an input parameter. The energy change for each candidate is calculated with:
\begin{align}
    p_{n+1} = p_n \left(1 - \frac{1}{3}\nabla\cdot\vec{u} \right) \: .
\end{align}

Since the divergence of the wind velocity, $\nabla\cdot\vec{u}$, is needed for the calculation, the \codeStyle{AdvectionField} class implements these values with a dedicated method. Adiabatic cooling is currently available for all implemented advection fields.

%% file: 03d_interactions.tex
\subsection{Radiation and Particle Interactions}
\label{ssec:interactions}

\begin{figure}[ht]
    \centering
    \includegraphics[width=0.9\textwidth]{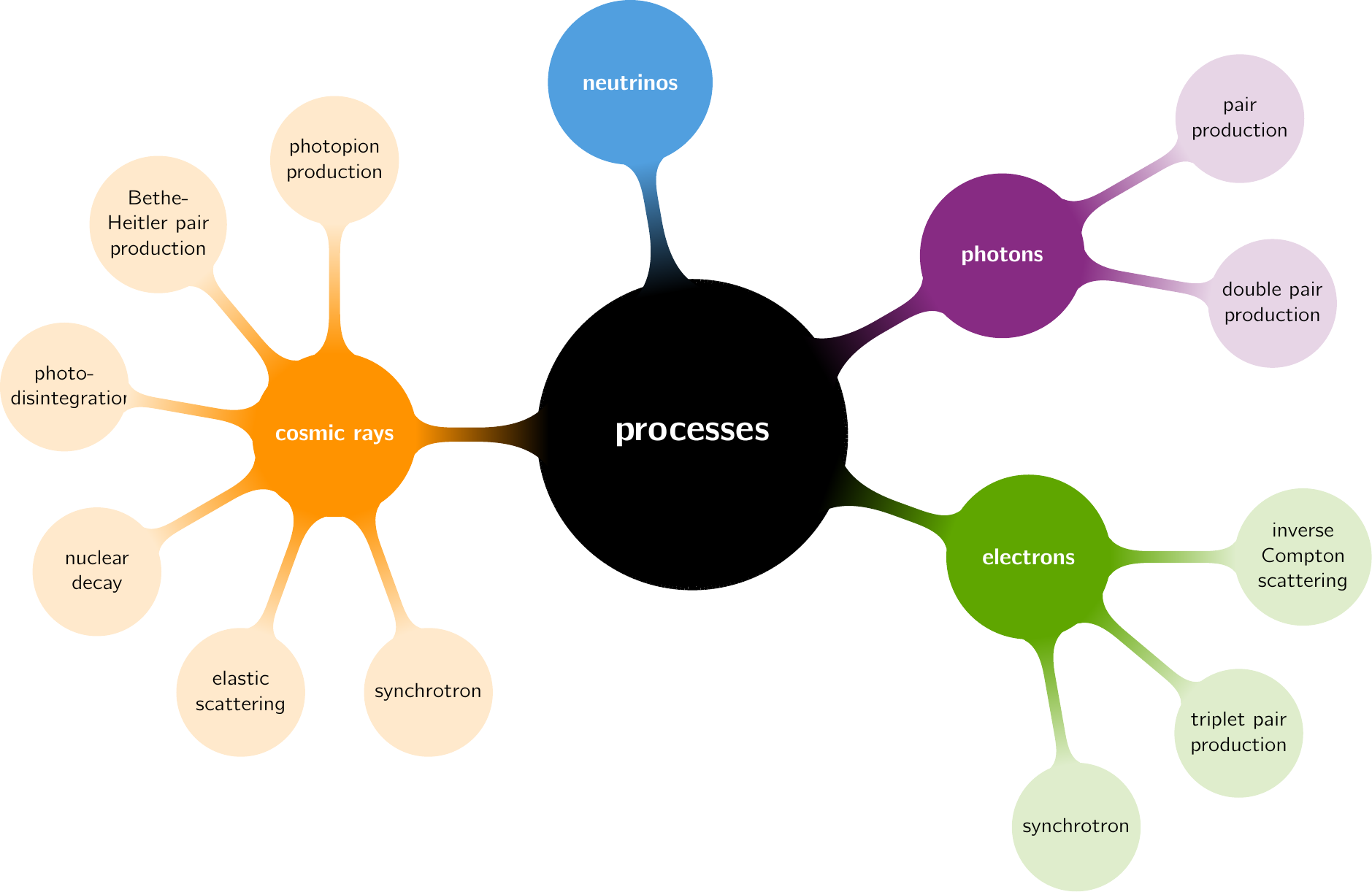}
    \caption{Schematic view of the interactions and energy-loss processes implemented in CRPropa~3.2 for different types of particles.}
    \label{fig:interactions}
\end{figure}

In this new version of CRPropa we introduced several features, adding new capabilities to the code and improving the way in which particle interactions are handled. In section~\ref{sssec:photonChannels} we describe the newly-implemented channels for photon production. In section~\ref{sssec:cascades} we present the new treatment of electromagnetic processes. Therefore, we can now model the photonuclear and electromagnetic interactions relevant for high-energy particle propagation, as shown in Fig.~\ref{fig:interactions}.
Among the technical developments, we describe the implementation of a thinning algorithm for efficient simulations of electromagnetic cascades in section~\ref{sssec:thinning}, as well as the improvements in the treatment of photopion production, in section~\ref{sssec:photopion}. Finally, section~\ref{sssec:customPhoton} contains a description of user-defined photon fields.

\subsubsection{Improvements in the Treatment of Photopion Production}
\label{sssec:photopion}

The treatment of photopion production in CRPropa is done with SOPHIA~\cite{muecke2000a}.
%
%
%
%
%
%
%
A direct CRPropa--SOPHIA interface has been implemented as a feature of the
\codeStyle{PhotoPionProduction} module, for the user to access. In this way, the SOPHIA particle interaction algorithm~\cite{muecke2000a} can be directly invoked to simulate the interaction of an arbitrary primary high-energy nucleon with a background photon. Here the primary particle must either be a proton or a neutron, since these are the only particles which SOPHIA accepts. In the case of nuclei, a re-scaling is performed based on the number of nucleons of each species (protons and neutrons), as done in previous versions of CRPropa~\cite{Kampert:2012fi, Batista:2016yrx}. Note that while this approximation is sufficient for the purposes of intergalactic cosmic-ray propagation, for other applications such as interactions in sources it may affect the flux of secondary particles~\cite{morejon2019a}.

\subsubsection{Update of the Photon Production Channels}
\label{sssec:photonChannels}

During UHECR propagation other messengers like photons, electrons, and neutrinos are produced in interactions with cosmic background radiation. Dominant production channels of these particles such as photopion production, electron pair production (Bethe-Heitler), photodisintegration, and nuclear decay were already implemented in the previous versions of CRPropa~\cite{Batista:2016yrx, Kampert:2012fi}. However, in these versions only the hadron, electron, and neutrino emission was considered; photon emission was only implemented for photopion production, which is by far the dominant channel for UHECR propagation over cosmological distances. In the release presented here we account for photons emitted in photodisintegration and nuclear decay. We also add a new interaction, the elastic scattering of cosmic-ray nuclei off background photons, which is yet another production channel for high-energy photons. In the following, the implementation of the new photon channels from Ref.~\cite{Heiter:2017cev} is summarised.

\paragraph{Photodisintegration.}
Photodisintegration is a nuclear process in which a high-energy cosmic-ray nucleus absorbs a photon from the cosmic background field, enters an excited state, and immediately decays by emitting one or more subatomic particles. CRPropa uses the TALYS~1.8~\cite{koning2005talys,TALYS1.8} nuclear code to simulate the interaction rates and branching ratios for all isotopes with $Z \leq 26$, $12 \leq A \leq 56$ and lifetimes $\tau > 2$\,s, as described in \cite{Kampert:2012fi,Batista:2016yrx}. After the interaction, the remaining nucleus can be in an excited state resulting in the emission of multiple high-energy photons depending on the number of nuclear-level transitions within the nucleus. To compute the photon emission by photodisintegration, TALYS 1.8 is again used for all nuclear fragments with $A \geq 7$. For nuclei with $A < 7$, currently no photon emission data is available. The photon emission ratios represent the relative emission abundance of photons with specific energies, depending on the nuclear-level scheme of the remaining nucleus, for all photodisintegration channels with the same initial-remaining nucleus combination. In the case of photodisintegration, the combination of initial and remaining nuclei is used to select the corresponding channel for photon emission.
Using tabulated energy-dependent emission probabilities of the photon channels, the photon energy is generated randomly and the photon then boosted to the observer frame.
All generated particles in the photodisintegration event are considered for energy conservation.

\paragraph{Nuclear decay.}
Photodisintegration may produce radioactive nuclei that subsequently suffer $\alpha$- or $\beta^{\pm}$-decays. The daughter nuclei are then generally left behind in an excited state from which they will decay by emitting one or several gamma rays, depending on their respective nuclear-level scheme.
The $\alpha$- and $\beta^{\pm}$-decays are implemented as described in \cite{Kampert:2012fi}, using the nuclear data of the NuDat~2.6~\cite{NuDat} database for all isotopes with $Z \leq 26$ and $A \leq 56$. For the implementation of the $\gamma$-decay, the NuDat data of the recorded photon energies and corresponding intensities are used to calculate photon emission probabilities for each discrete photon energy within 124 $\gamma$-decay channels. The number of photons emitted by each nucleus can be larger than one, depending on the excitation energy at which the produced nucleus is formed and the specific properties of the nuclear level scheme. But it can also be smaller than one because not all preceding decays will leave the nucleus in an excited state. Photon emissions caused by other processes listed in the NuDat database, like positron annihilation or X-ray emission in electron shell processes are not considered because they do not apply to the fully ionised cosmic rays considered here. 
As before, the created photons are boosted to the observer frame and are considered, for energy conservation.

\paragraph{Elastic Scattering.}
Photons from the cosmic background radiation can scatter elastically off cosmic-ray nuclei. In this process, the nuclei transfer part of their energy to the photon, similarly to inverse Compton scattering. The interaction length for this process is calculated for all isotopes with $Z \leq 26$, $12 \leq A \leq 56$ and lifetimes $\tau > 2$~s using the cross-section data simulated with TALYS~1.8~\cite{koning2005talys,TALYS1.8}. In comparison to photodisintegration, this process is expected to be sub-dominant. Therefore, an approximation is used to reduce the amount of tabulated data and to save internal memory~\cite{Heiter:2017cev}. The individual interaction rates of the isotopes are obtained from the tabulated average interaction rate via $\lambda^{-1}_{Z,A} \approx (ZN/A)\,\lambda_{\mathrm{avg}}^{-1}$, where $Z,N,A$ correspond to the nuclear charge, neutron, and mass number, respectively. A similar scaling approach is applied to the differential interaction rate $\mathrm{d}\lambda^{-1}_{\mathrm{avg}}/\mathrm{d}\varepsilon$ which is used to sample the energy of the interacting background photon $\varepsilon$ in the reference frame. When an interaction occurs, the sampled background photon is boosted to the observer frame and considered for energy conservation. The uncertainty on the photon energy introduced by the applied scaling method is small compared to the general simulation uncertainties~\cite{Heiter:2017cev}.

\subsubsection{Implementation of Electromagnetic Cascades}
\label{sssec:cascades}

In the previous version of CRPropa, propagation of electromagnetic particles ($\gamma$, $e^{\pm}$) was performed separately using the stand-alone cascading codes EleCa~\cite{settimo2015propagation} and DINT~\cite{lee1998propagation}. To enable simultaneous electromagnetic and hadronic propagation, interaction modules for the dominant electron and photon processes are implemented and can be added directly to the CRPropa simulation chain. These processes are: pair production (Breit-Wheeler), double pair production, triplet pair production, and inverse Compton scattering. In addition, an interaction module for synchrotron radiation was added to the code. The implementation of the electromagnetic interaction modules is based on the EleCa code as described in~\cite{Heiter:2017cev}, while several improvements were made, which have an impact on the generated secondary particle spectra. 
For all interaction modules (\codeStyle{EMPairProduction}, \codeStyle{EMInverseComptonScattering}, \codeStyle{EMPairDoubleProduction}, \codeStyle{EMTripletPairProduction}), except for synchrotron radiation (\codeStyle{SynchrotronRadiation}), the individual interaction rates $\lambda^{-1}(E)$ and differential interaction rates $\mathrm{d}\lambda^{-1}/\mathrm{d}s(s,E)$ are tabulated using the corresponding cross-section data. In case of an interaction via pair production or inverse Compton scattering, the differential interaction rate is used to sample the squared centre-of-mass energy $s$. Together with the differential cross section $\mathrm{d}\sigma(s,x)/\mathrm{d}x$, the energy ratio $x$ between the initial and remaining particle is sampled\footnote{Note that the current implementation of the sampling in CRPropa~3.2 is not affected by issues present in previous version, discussed in Ref.~\cite{Kalashev:2022cja}.}. Using energy conservation the energies of all particles produced within the interaction are determined. For double and triplet pair production common approximations are used to determine the particle energy after interaction~\cite{Heiter:2017cev}. The \codeStyle{SynchrotronRadiation} module is implemented as continuous energy-loss process because high-energy charged particles with low masses, like electrons, emit a huge amount of synchrotron photons while moving through magnetic fields. The  parametrisation of Ref.~\cite{jackson1999classical} is used to sample the amount of energy emitted via synchrotron emission within the propagation step, while synchrotron photons with energies according to the synchrotron emission spectrum are generated until the total energy expected to be lost within the propagation step is reached. The last synchrotron photon is accepted randomly to ensure energy conservation on average.

\subsubsection{Thinning}
\label{sssec:thinning}

The propagation of high-energy photons in intergalactic space is an inherently difficult problem due to the growing number of secondary particles produced in electromagnetic cascades. 
For this reason, we have introduced a weighted sampling procedure in CRPropa, largely inspired by the ELMAG code~\cite{Kachelriess:2011bi, Blytt:2019xad}, and following the implementation of~\cite{AlvesBatista:2016urk}. In essence, whenever a secondary particle is produced in a process, a random number $r$ is drawn to determine whether or not the particle is accepted.  Acceptance/rejection depends on the fraction $f$ of the energy of the primary retained by the secondary. Therefore, by choosing a suitable thinning parameter $\eta$, only particles for which the condition $r < (1 - f)^\eta$ is satisfied are tracked further, with their weights adjusted accordingly: $w = w_0 f^{-\eta}$, wherein $w_0$ refers to the original weight of the primary particle before the interaction. Here, the parameter $\eta \in [0, 1]$ controls whether no thinning is applied ($\eta = 0$), or whether it is maximal ($\eta = 1$). The optimal value of $\eta$ is a compromise between performance and accuracy for a sample of limited size. Therefore, $\eta$ depends on the particular problem at hand. 

For the propagation of cascades initiated by gamma rays with a few TeV for sources at redshifts $\lesssim 0.20$, the typical performance improvement achieved by a maximal thinning can be of up to two orders of magnitude compared to the case where no thinning is used. For cascades started by higher-energy particles with EeV energies, this gain can be even higher, around five orders of magnitude.

\subsubsection{Introduction of Custom Photon Fields}
\label{sssec:customPhoton}

Prior to this work, CRPropa only provided a pre-defined set of photon fields relevant for particle propagation at (inter-)Galactic distances. We now introduce a new handling of photon fields which, next to several internal optimisations, enables the user to add their own custom photon fields to CRPropa.

On-the-fly calculations of, e.g., mean free paths of a particle within a photon field, would be too costly from a computational point of view, so that these quantities are provided as pre-computed tabular data files. All these files are automatically downloaded upon installation and are stored in a designated location from where it is accessed by the respective CRPropa module. For reference, all relevant scripts to generate these files are accessible through a separate repository.\footnote{\url{https://github.com/CRPropa/CRPropa3-data}} In particular, the redshift scaling factor (RSF) is a computational shortcut to represent the evolution of the number density of a photon field with redshift. In the case of the cosmic microwave background (CMB), which evolves adiabatically, the evolution of the number density is a trivial scaling with redshift of the form $(1 + z)^2$. In the more general case of fields that do not evolve trivially, the scaling factor is defined as the ratio of the comoving photon number density at some given redshift, $z$, over the initial, i.e.\ the maximal comoving photon number density at $z=0$, obtained through:
\begin{equation}
    {\rm{RSF}}(z)=\int\limits_{\varepsilon_{\min}}^{\varepsilon_{\max}}
    \frac{{\rm{d}}n_\gamma}{{\rm{d}}\varepsilon_\gamma}(z) {\rm{d}}\varepsilon_\gamma \, \left[ \int\limits_{\varepsilon_{\min}}^{\varepsilon_{\max}} \frac{{\rm{d}}n_\gamma}{{\rm{d}} \varepsilon_\gamma} (z=0) {\rm{d}}\varepsilon_\gamma \right]^{-1} \:.
    \label{eq:RSF}
\end{equation}
Here, $\varepsilon_{\min}$ and $\varepsilon_{\max}$ are the minimal and maximal photon energies between which the photon field is defined.

In CRPropa~3.2, all previous implementations of photon fields were replaced by their corresponding class representations. This approach not only allows us to streamline the software architecture of target fields with the magnetic field and matter density classes but also centralises data related to photon fields for much smarter data handling. Furthermore, the introduced changes provide a general platform for the user to customise photon fields. This particularly includes the ex-situ construction of photon field objects in a brick-like way as it is also realised by, e.g., the magnetic field classes.

One newly-implemented class of photon field is the \codeStyle{BlackbodyPhotonField}. The blackbody photon field is further specialised by \codeStyle{CMB}, which is now represented by an object (as opposed to the previous CMB enumerator object), inheriting all properties of the parent \codeStyle{BlackbodyPhotonField}. In the case of a blackbody photon field, the photon number density is described by well-behaved analytical expressions. However, the density of other photon fields, such as current extragalactic background light models, may have been acquired (semi-)empirically and thus may not have an analytical representation. Such photon fields --- in this work and in the code referred to as extragalactic background light (EBL) or intergalactic radiation background (IRB) --- are handled by the \codeStyle{TabularPhotonField} class. In principle, this class allows for the initialisation of any tabulated photon field data.

%% file: 03e_acceleration.tex
\subsection{Acceleration}
\label{ssec:acceleration}

The new version of CRPropa adds an important ingredient to build complete models of high-energy emission: particle acceleration. It is now possible to simulate the acceleration of particles through their scattering off scattering centres representing, e.g., irregularities in the magnetic field. This is a generalisation of the implementation used in~\cite{Winchen2018}.

The modules scatter the particles into a random direction in the rest-frame of a scattering centre moving with normalised velocity $\beta = v/c$ in the laboratory frame after propagating a distance $d$ randomly chosen according to an exponential distribution with mean free path $\lambda$.
Acceleration is achieved depending on the properties of the movement of the scattering centres via the second- or first-order Fermi mechanism.

Acceleration via second-order Fermi mechanism is modelled by assuming isotropic movements of the cosmic rays and scattering centres in the rest frame of the accelerating region. First-order Fermi acceleration is achieved by assuming a directed flow of scattering centres in two regions with different velocities.
The functionality is separated into distinct classes accounting for effects of geometry, modification of the step length, and properties of the velocity distribution of the scattering centres. This way, the user can study the impact of each individual effect separately.

%% file: 04_Examples.tex
\section{Example Applications Focussing on New Features}
\label{sec:examples}

\subsection{Galactic Propagation}
\label{ssec:GalPropExample}

In galactic environments the motion of cosmic rays often can be described by diffusion. As shown in section~\ref{ssec:galactic}, CRPropa solves the transport equation within the \codeStyle{DiffusionSDE} module.
The impact of the magnetic field on the Galactic centre is shown in a sample simulation\footnote{The simulation setup of the following example can be found in the CRPropa documentation at https://crpropa.github.io/CRPropa3/} for Galactic cosmic rays.

We take the \codeStyle{DiffusionSDE} module with the default values for the step size and the ratio $\epsilon = {\kappa_\perp}/{\kappa_\parallel}$ of perpendicular and parallel diffusion. The energy dependence of the diffusion coefficient is:
\begin{equation}
\kappa_\parallel = \kappa_0  \left(\dfrac{E}{4 \, \textrm{GeV}}\right)^\frac{1}{3} \: .
\end{equation}
For the magnetic field we only use the \codeStyle{JF12Solenoidal}~\cite{Kleimann2019} in the first run without the outer transition and without the turbulent component. In the second run we use the superposition of the \codeStyle{JF12Solenoidal} field with the inter-cloud component of \cite{Guenduez2020} in the \codeStyle{CMZField} module. Additionally, more detailed structures of the \codeStyle{CMZField} can be neglected for this type of analysis on Galactic scales.

The source density of the cosmic rays follows the pulsar distribution of \cite{Faucher_Giguere_2006} with the parametrisation of the radial distribution of \cite{Blasi_2012} (\codeStyle{SourcePulsarDistribution}). At this point we only take protons with a fixed energy of $ E = 10 \, \textrm{TeV}$, for simplification. In our setup the simulation volume is limited to a cylinder with the height of $z = \pm 2~\textrm{kpc}$ over the Galactic plane and a Galactocentric radius of $r_\mathrm{gc} = 20~\textrm{kpc}$.
To get a stationary solution of the transport equation we use the weighting approach presented in \cite{Merten:2017mgk}. Therefore, we observe the cosmic rays with a \codeStyle{ObserverTimeEvolution} with $n_{\textrm{step}} = 100$ and $\Delta t = 5$\,kpc/$c$.

\begin{figure}[ht!]
    \centering
    \includegraphics[width=\textwidth]{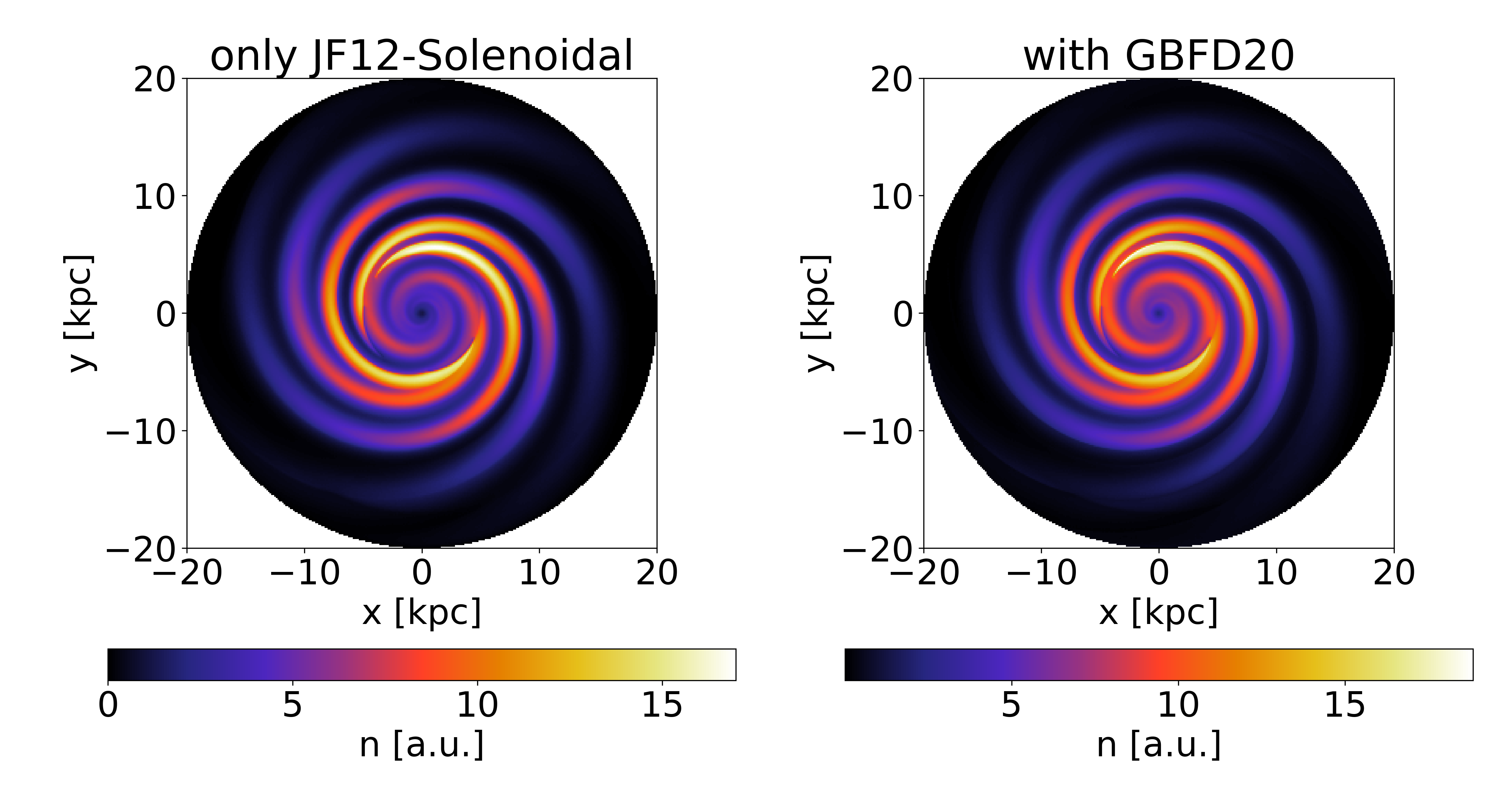}
    \caption{Face-on view of the cosmic-ray density for a simulation with the \codeStyle{JF12Solenoidal} field \cite{Kleimann2019} (left panel) and the superposition of \codeStyle{JF12Solenoidal} and the GBFD20 field \cite{Guenduez2020} (right panel). Here, only particles inside the Galactic disc ($-300 ~\textrm{pc}\leq z \leq 300~ \textrm{pc}$) are shown. The density $n$ is given in pseudo-particles per bin. }
    \label{fig:ExampleAbsVal}
\end{figure}


Figure~\ref{fig:ExampleAbsVal} shows the face-on views of the stacked cosmic-ray density with the \codeStyle{JF12Solenoidal} field~\cite{Kleimann2019} and the superposition of \codeStyle{JF12Solenoidal} and the GBFD20 field~\cite{Guenduez2020}.
It shows that the density mainly follows the spiral structure of the magnetic field. The differences in the Galactic centre region are also visible. 

\subsection{Electromagnetic Cascades and Thinning}

With CRPropa 3.2, it is now possible to perform the propagation of electromagnetic cascades in the intergalactic medium very efficiently. 
The applications to gamma-ray astrophysics are numerous. As an example, we study multi-TeV emission from the extreme blazar 1ES~0229+200, an archetypal object commonly used to constrain intergalactic magnetic fields (see~\cite{alvesbatista2021a} for a review on the topic). 

We consider five magnetic-field strengths: $B = 10^{-14} \; \text{G}$, $10^{-15}$, $10^{-16} \; \text{G}$, $10^{-17} \; \text{G}$, and 0. We fix the coherence length to $L_B = 1 \; \text{Mpc}$ and employ the turbulent magnetic fields described in section~\ref{sssec:turbulentFields}. We also set the thinning parameter to $\eta=1$ (see section~\ref{sssec:thinning}).

\begin{figure}[ht!]
    \centering
    \includegraphics[width=0.8\textwidth]{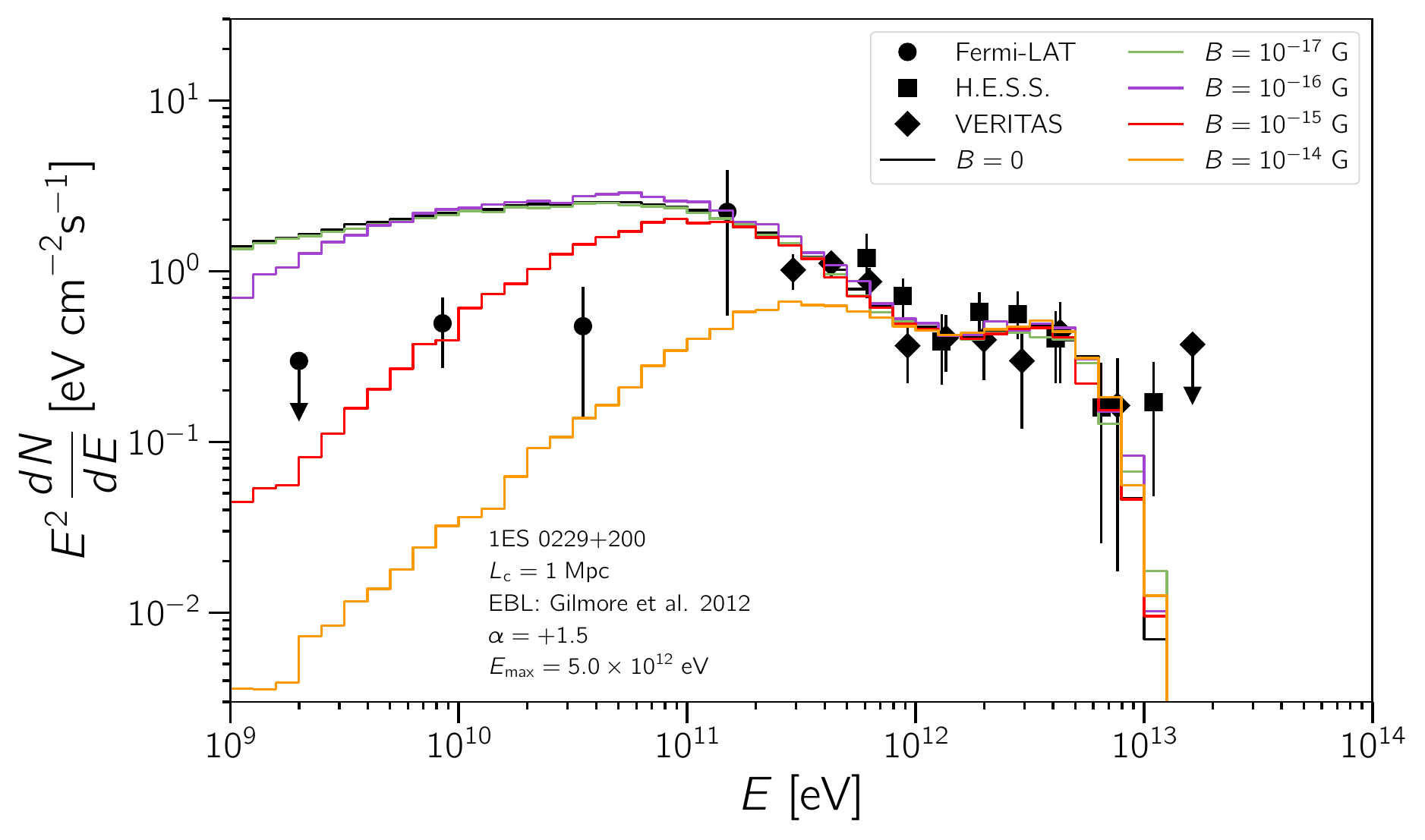}
    \caption{Gamma-ray flux arriving at Earth from the blazar 1ES~0229+200. The object is assumed to emit gamma rays directly towards Earth, with a power-law index $\alpha=1.5$ and an exponential cut-off at $E_\text{max} = 5 \times 10^{12} \; \text{eV}$. The magnetic field is assumed to have a coherence length $L_B = 1 \; \text{Mpc}$ and strengths $10^{-17} \; \text{G}$ (green), $10^{-16} \; \text{G}$ (purple), $10^{-15} \; \text{G}$ (red), and $10^{-14} \; \text{G}$ (orange). The scenario without magnetic fields is represented by the black line. For reference, Fermi-LAT (circles) and H.E.S.S.~\cite{Aharonian:2007wc} (squares) measurements are also shown.}
    \label{fig:cascade_spec}
\end{figure}

\begin{figure}[ht!]
    \centering
    \includegraphics[width=0.49\textwidth]{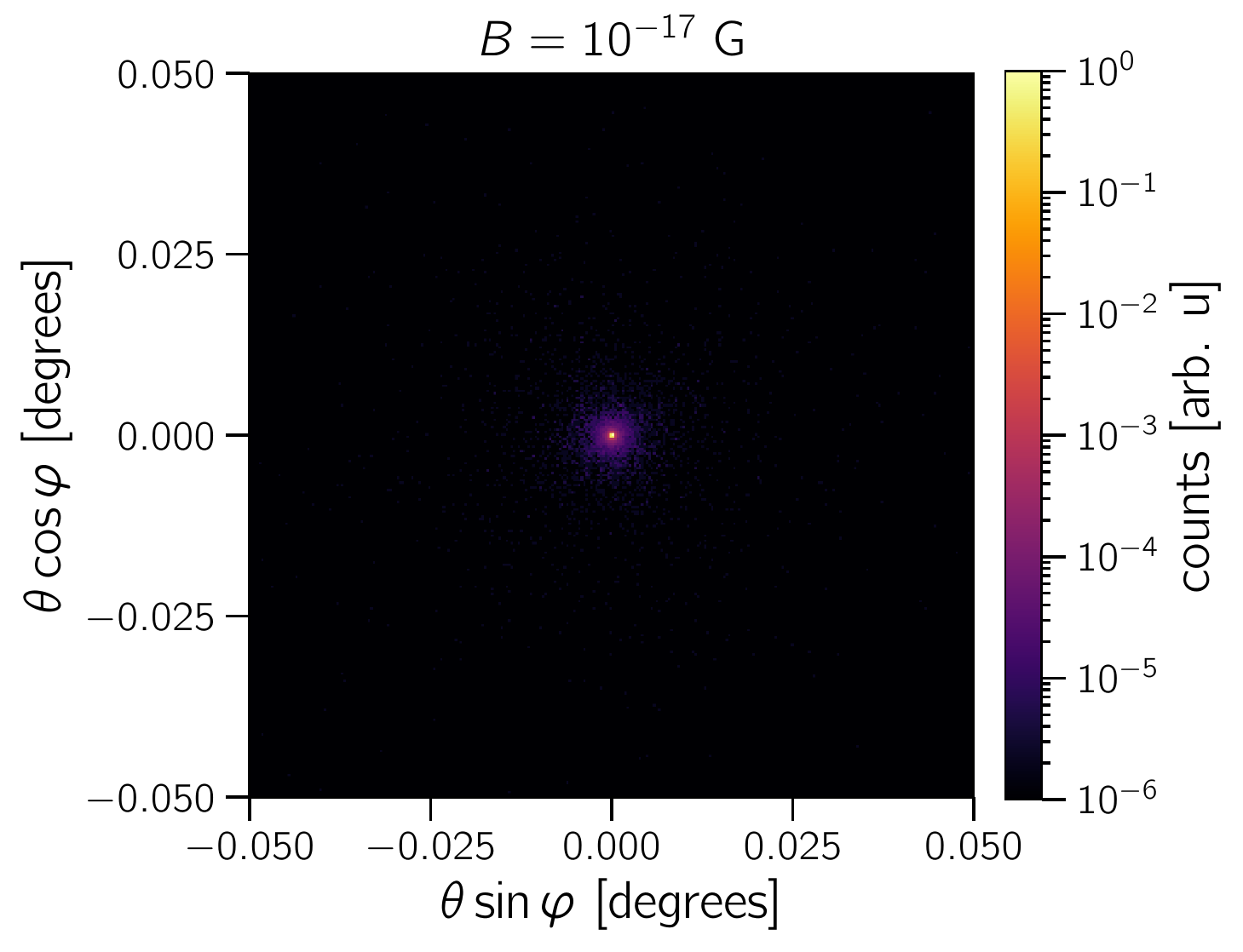}
    \includegraphics[width=0.49\textwidth]{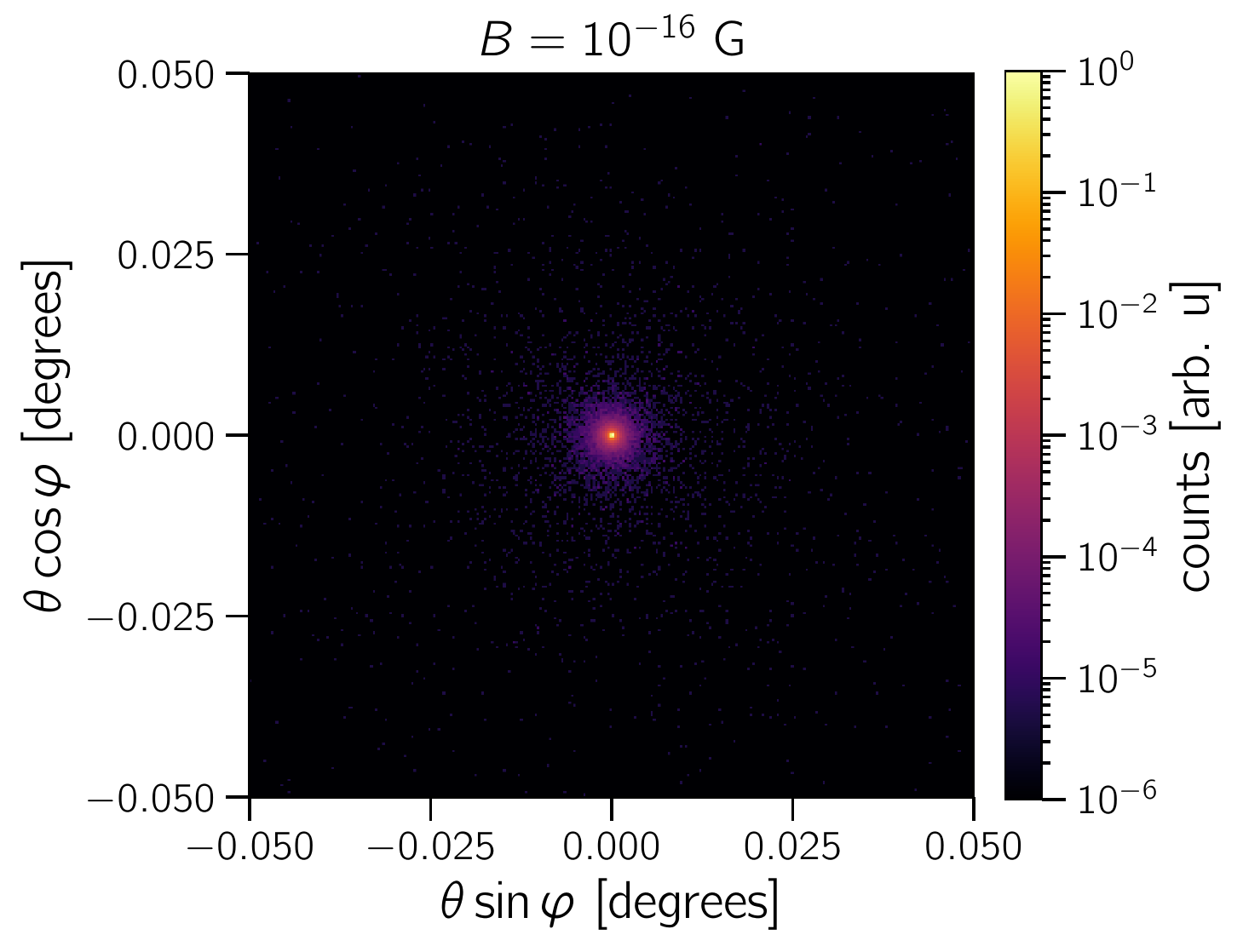}
    \includegraphics[width=0.49\textwidth]{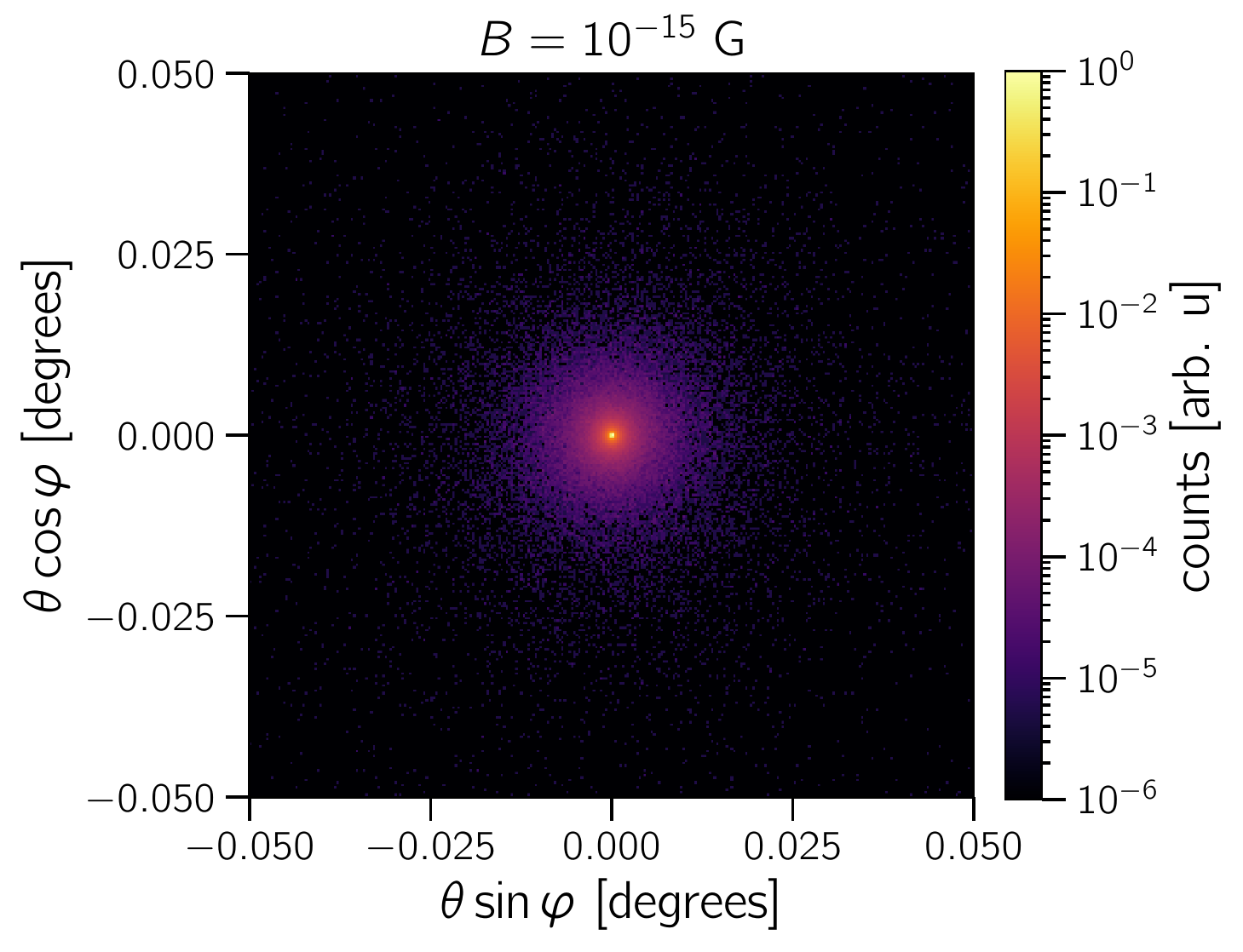}
    \includegraphics[width=0.49\textwidth]{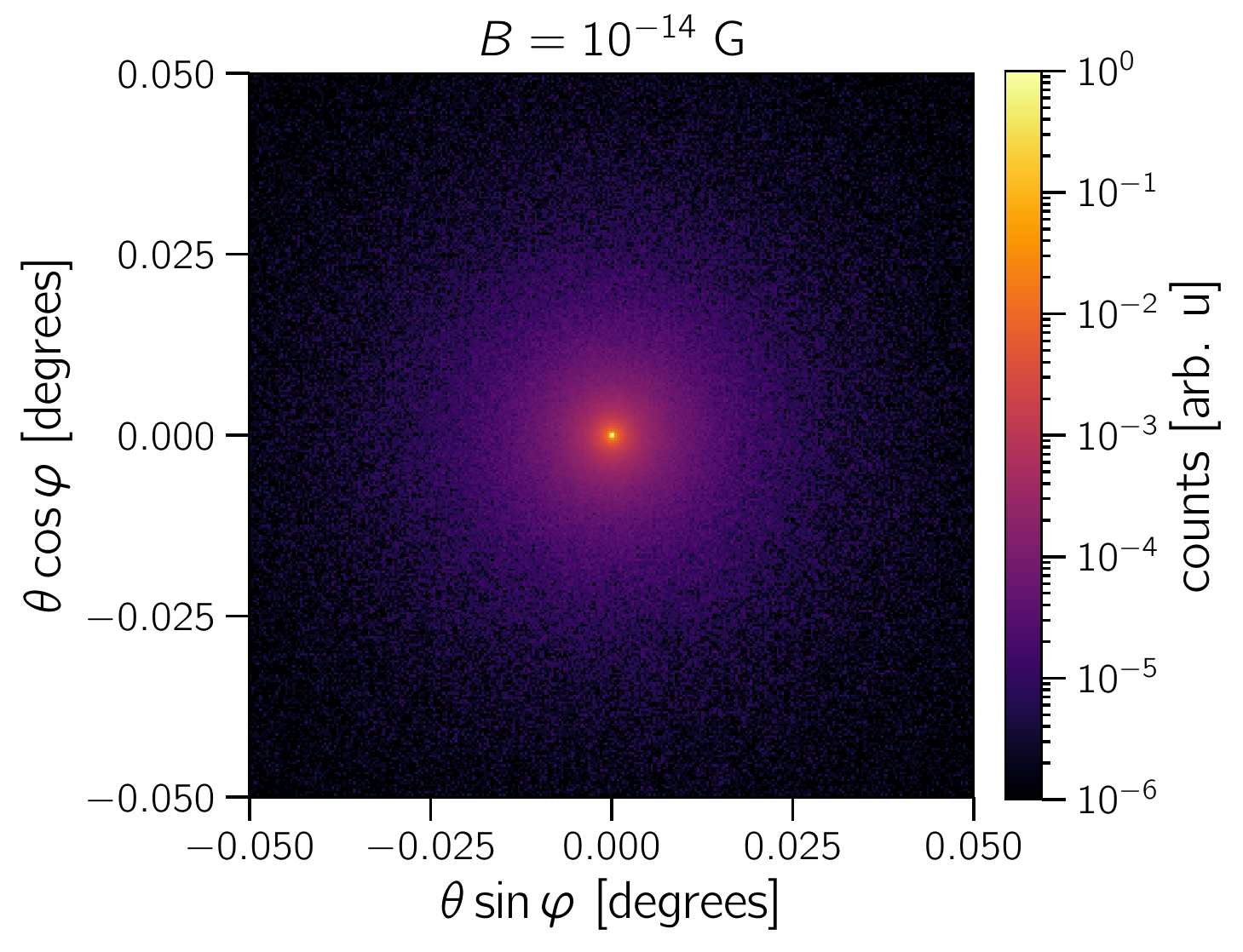}
    \caption{Arrival directions of the gamma rays propagated from 1ES~0229+200 for a magnetic field strengths of $10^{-17} \; \text{G}$ (upper left), $10^{-16} \; \text{G}$ (upper right), $10^{-15} \; \text{G}$ (lower left), and $10^{-14} \; \text{G}$ (lower right panel). The parameters chosen for this example are the same as in Fig.~\ref{fig:cascade_spec}.}
    \label{fig:cascade_haloes}
\end{figure}

In Fig.~\ref{fig:cascade_spec} we plot the energy spectrum of the gamma rays measured at Earth. As the magnetic-field strength increases, a suppression starts to become visible in the gamma-ray flux as energy decreases. This can be better understood by analysing Fig.~\ref{fig:cascade_haloes}, which shows the arrival directions of the gamma rays for four magnetic-field strengths ($10^{-17}$, $10^{-16}$, $10^{-15}$, and $10^{-14} \; \text{G}$). The larger angular spread for stronger magnetic fields leads to a decrease of the measured point-like flux, leading to the suppression observed in Fig.~\ref{fig:cascade_spec}.

The electromagnetic cascades modules of CRPropa have been successfully used in a number of works similar to this example, including Fermi-LAT constraints on IGMFs~\cite{Biteau:2018tmv}, the sensitivity of the Cherenkov Telescope Array (CTA) for constraining IGMFs~\cite{Abdalla:2020gea}, or multimessenger IGMF constraints~\cite{AlvesBatista:2020oio}.

The advantage of the native electromagnetic cascade modules, in contrast to the previous approach using DINT/EleCa, is that it allows us to capitalise on one of CRPropa's strengths: its modularity. For instance, in Ref.~\cite{AlvesBatista:2019ipr} a plugin
was developed to investigate the effect of plasma instabilities on cascades which can be easily used together with other CRPropa modules.


\subsection{Custom Photon Fields}

The investigation of several local astrophysical environments brings up the need to account for the local photon fields to realistically model the interactions. CRPropa now offers the possibility to use custom, user-defined photon fields, as presented in section\,\ref{sssec:customPhoton}. We illustrate the usage of this new feature considering a proton beam with a fixed energy of $E=10^{19.5}\,\mathrm{eV}$ injected into three different isotropic photon targets. First, we consider the case of the CMB, where the Greisen–Zatsepin–Kuzmin (GZK) effect is expected to set in around these energies. Second, for an analytically-described field, we consider the radiation field due to a blackbody with temperature $T=100\,\mathrm{K}$. Finally, to demonstrate the possibility of using arbitrarily-shaped fields, we use tabulated data corresponding to the EBL model by Gilmore~\textit{et al.}~\cite{gilmore2012a}. For a consistent comparison, the optical depth is fixed to $\tau = 1$ in all cases studied. In this simple one-dimensional simulation we inject $10^{4}$ primary CRs and include only photopion production.

\begin{figure}[!ht]
    \centering
    \includegraphics[width=0.7\textwidth]{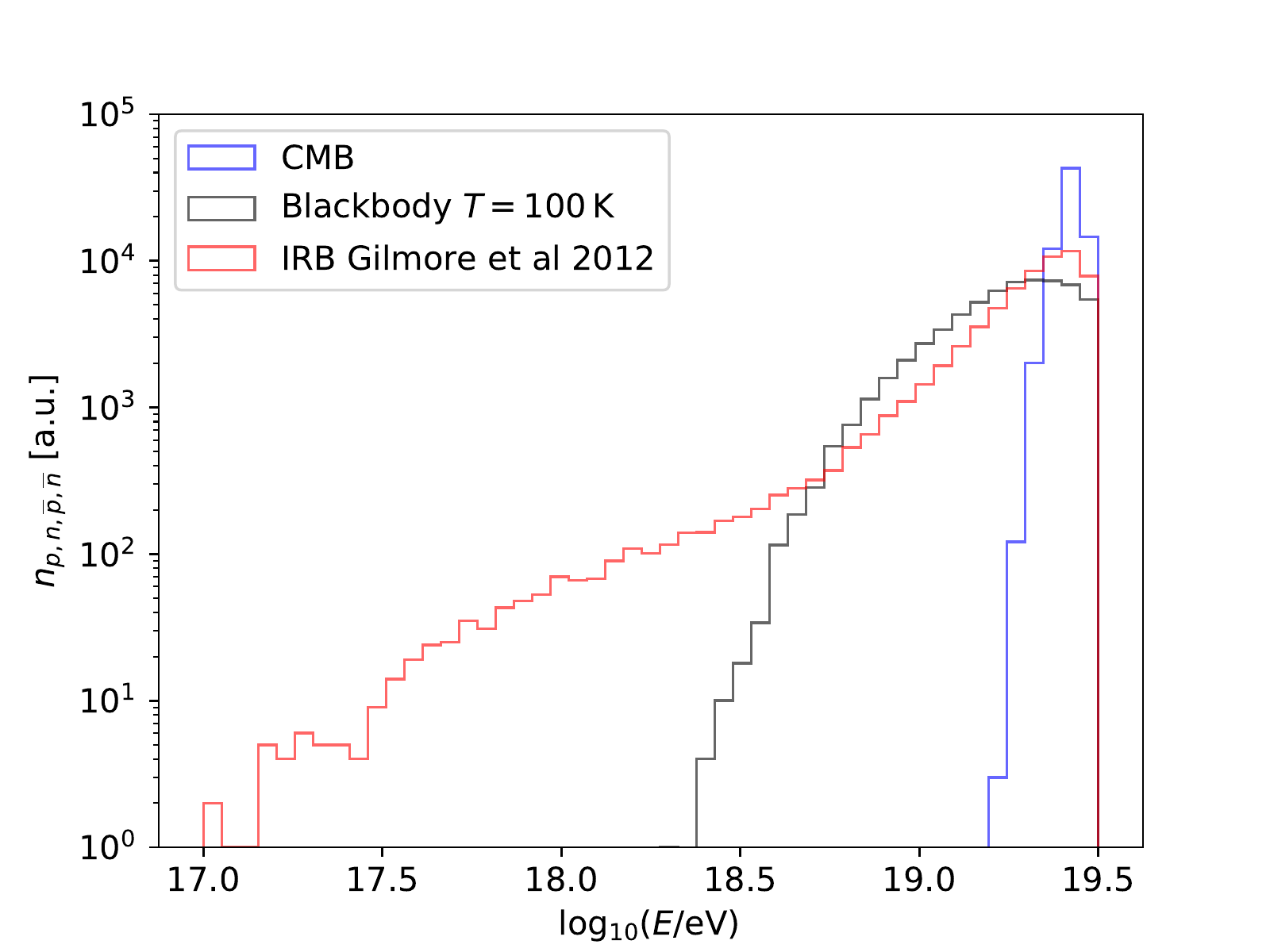}
    \caption{Histogram of the observed protons, neutrons, and their antiparticles after the interaction of $10^{19.5}\,\mathrm{eV}$ protons with the CMB, with the EBL model by Gilmore~\textit{et al.}~\cite{gilmore2012a}, and with a blackbody with temperature of $100\,\mathrm{K}$.}
    \label{fig:ExampleCusField:nucleons}
\end{figure}

Figure \ref{fig:ExampleCusField:nucleons} shows the number of particles per logarithmic energy bin for all observed nucleons that were detected (mainly the protons that suffered energy losses and neutrons produced via photopion production) after the primaries propagated through the three different target photon fields. While the distribution in the case of the least energetic field (the CMB) is mostly concentrated slightly below the injection energy, for the higher temperature blackbody and especially for the infrared background, much broader distributions are observed, extending many orders of magnitude down in energy. 

\begin{figure}[!ht]
     \centering
     \includegraphics[width=1.0\textwidth]{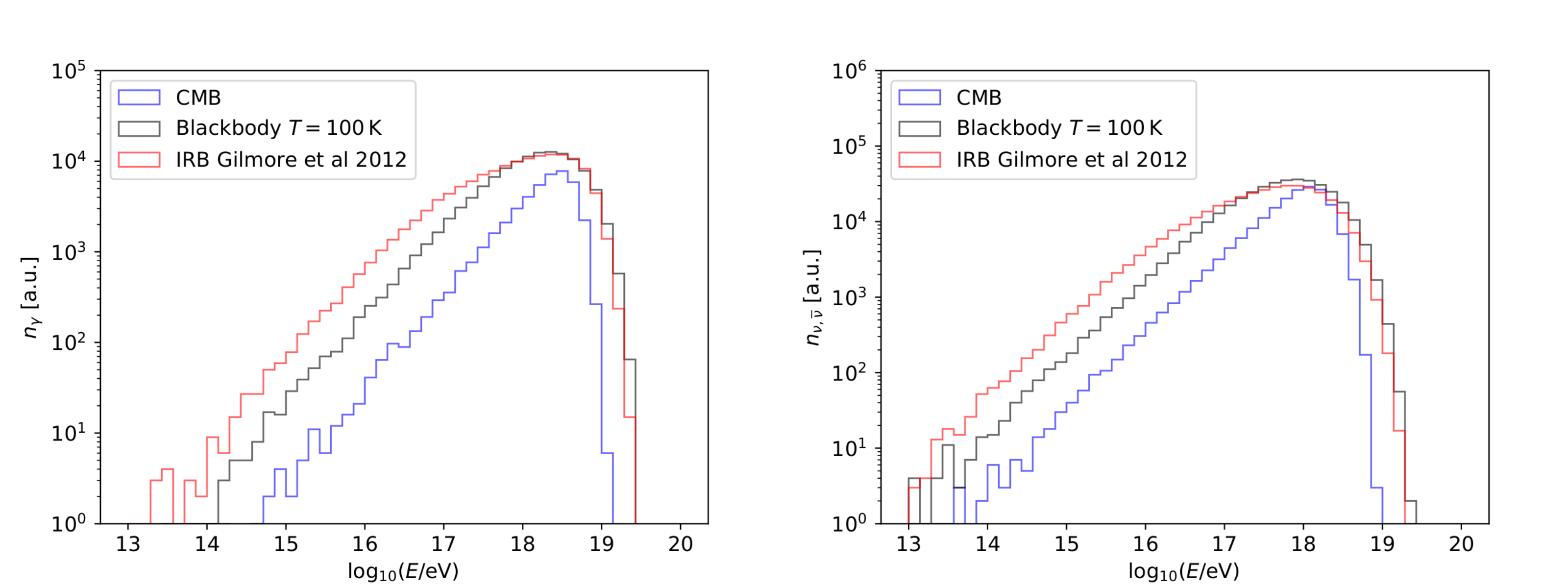}
     \caption{Histogram of the observed secondary photons (left) and neutrinos and their antiparticles (right panel) resulting from the decay of the pions produced through the interaction of $10^{19.5}\,\mathrm{eV}$ protons with the CMB, with the EBL~\cite{gilmore2012a}, and with a blackbody with temperature of $100\,\mathrm{K}$.}
    \label{fig:ExampleCusField:gn}
\end{figure}

The number of secondary photons and neutrinos resulting from the decay of the produced pions are shown in Fig.~\ref{fig:ExampleCusField:gn}, illustrating a typical multimessenger information that can be obtained. Their distributions roughly follow the same order as the total energy of the photon field, but do not vary as much in shape.

In a similar manner, these custom fields can now be used in CRPropa to model a variety of realistic scenarios such as, for example, the local environment of an accretion disc in an active galaxy~\cite{hoerbe2020a} or neutrino and gamma-ray signatures of PeVatrons~\cite{rodriguezramirez2019c}.

%% file: 05_Summary.tex
\section{Summary and Outlook}
\label{sec:summary}

With the recent advances in astroparticle physics and high-energy astrophysics, it is important to understand the production and propagation of cosmic rays, photons, and neutrinos within a single self-consistent framework. To obtain a global picture of these messengers, advanced computational tools are required, enabling the exploration of the multi-dimensional parameter space relevant for interpreting the available observational data.

Here we introduced the new version of the publicly available software for high-energy particle propagation --- CRPropa~3.2. It inherits all features from the previous version and extends its capabilities to tackle several other astrophysical problems. New physics features include improved treatment and implementation of new models of magnetic fields, a native treatment of electromagnetic cascades, updated channels for photon production, modules for first- and second-order Fermi acceleration, support for user-defined photon fields, and Galactic gas density models. On the technical side, this version implements additional propagation modes using the Boris push algorithm, an ensemble-averaged treatment of propagation using stochastic differential equations, enhanced grid interpolation methods, and faster three- and four-dimensional propagation with the targeting technique.

In the near future, CRPropa will be extended to become an integrated simulation framework for the production and propagation of (ultra-)high-energy particles from inside astrophysical sources, through intergalactic space and the Galaxy, down to the Solar System, spanning cosmic-ray energies from GeV up to ZeV. The code will be able to handle all particles ranging from photons and neutrinos up to the heaviest nuclei in a consistent way. Until then, the modular structure of CRPropa allows the user to tailor it to their needs and efficiently build the extensions necessary for their specific applications. These developments can now also be shared via plugins for the benefit of the community.

\bigskip
Information on how to download the code, an online documentation, and examples of applications  (including some of those presented in this work) can be found in {\url{https://crpropa.desy.de/}}. 